\newif{\ifarxiv}
\newif{\ifdraft}
\newif{\ifremarks}

\arxivtrue  
\remarkstrue  

\documentclass[11pt,a4paper]{article}
\ifdraft\overfullrule=5pt\fi
\pdfoutput=1
\usepackage[utf8]{inputenc}
\usepackage[left=2.5cm,right=2.5cm,top=2.8cm,bottom=2.8cm]{geometry}
\usepackage{graphicx}
\usepackage{amsmath,amssymb,mathtools}
\ifdraft\usepackage{showkeys}\fi 
\usepackage[bookmarks=true,ocgcolorlinks=true]{hyperref} 
\usepackage{cite}
\usepackage{xcolor}
\usepackage{booktabs}
\usepackage{multirow}
\usepackage{graphbox}
\usepackage[mathbold,autobold,greekcaps,greeklower]{mathfixs}
\renewcommand{\mathbf}{\mathbold}

\usepackage{lmodern}
\usepackage[T1]{fontenc}
\usepackage{microtype}
\usepackage{marvosym}

\usepackage{pifont}
\newcommand{\cmark}{\ding{51}}%
\newcommand{\xmark}{\ding{55}}%

\definecolor{green1}{HTML}{58ae3d}
\definecolor{cyan1}{HTML}{37cdaa}
\definecolor{blue1}{HTML}{5d7ac4}
\definecolor{red1}{HTML}{d0482a}
\definecolor{purple1}{HTML}{845ea8}
\definecolor{orange1}{HTML}{e07229}
\definecolor{magenta1}{HTML}{C45D7A}

\ifremarks
\newcommand{\remarks}[1]{{\renewcommand{\bfdefault}{b}{\color[RGB]{0,0,150}{\textbf{#1}}}}}

\fi
\providecommand{\remarks}[1]{\ignorespaces}

\allowdisplaybreaks
\usepackage[font=small,labelfont=bf,width=0.85\textwidth]{caption}

\providecommand{\hypersetup}[1]{}
\providecommand{\hyperref}[2][]{#2}
\providecommand{\texorpdfstring}[2]{#1}
\providecommand{\pdfbookmark}[3][]{}

\hypersetup{plainpages=false}
\hypersetup{pdfpagemode=UseOutlines}
\hypersetup{bookmarksnumbered=true}
\hypersetup{bookmarksopen=true}
\hypersetup{pdfstartview=FitH}
\hypersetup{citecolor=[rgb]{0 .4 0}}
\hypersetup{urlcolor=[rgb]{.4 0 0}}
\hypersetup{linkcolor=[rgb]{0 0 .4}}
\hypersetup{citebordercolor={.6 .9 .6}}
\hypersetup{urlbordercolor={.7 .8 1}}
\hypersetup{linkbordercolor={1 .7 .7}}
\hypersetup{pdfborder={0 0 .5}}

\makeatletter
\renewcommand*\l@section[2]{%
  \ifnum \c@tocdepth >\z@
    \addpenalty\@secpenalty
    \addvspace{0.7em \@plus\p@}%
    \setlength\@tempdima{1.5em}%
    \begingroup
      \parindent \z@ \rightskip \@pnumwidth
      \parfillskip -\@pnumwidth
      \leavevmode \bfseries
      \advance\leftskip\@tempdima
      \hskip -\leftskip
      #1\nobreak\hfil \nobreak\hb@xt@\@pnumwidth{\hss #2}\par
    \endgroup
  \fi}
\makeatother

\newcommand{\namedref}[2]{\hyperref[#2]{#1~\ref*{#2}}}
\newcommand{\secref}[1]{\namedref{Section}{#1}}

\newcommand{\appref}[1]{\namedref{Appendix}{#1}}

\newcommand{\tabref}[1]{\namedref{Table}{#1}}

\newcommand{\figref}[1]{\namedref{Figure}{#1}}

\makeatletter
\def\mr@ignsp#1 {\ifx\:#1\@empty\else #1\expandafter\mr@ignsp\fi}%
\newcommand{\multiref}[1]{\begingroup
\xdef\mr@no@sparg{\expandafter\mr@ignsp#1 \: }%
\def\mr@comma{}%
\@for\mr@refs:=\mr@no@sparg\do{\mr@comma\def\mr@comma{,\,}\ref{\mr@refs}}%
\endgroup}
\makeatother
\renewcommand{\eqref}[1]{(\multiref{#1})}

\makeatletter
\let\@myabstract\@empty
\let\@keywords\@empty
\let\@subject\@empty
\providecommand{\affiliation}[1]{\gdef\@affiliation{#1}}
\providecommand{\myabstract}[1]{\gdef\@myabstract{#1}}
\providecommand{\keywords}[1]{\gdef\@keywords{#1}}
\providecommand{\subject}[1]{\gdef\@subject{#1}}
\def\thetitle{\@title}
\def\theauthor{\@author}
\def\theaffiliation{\@affiliation}
\def\theabstract{\@myabstract}
\def\thesubject{\@subject}
\def\thedate{\@date}
\def\thekeywords{\@keywords}
\makeatother
\def\fillpdfdata{
\hypersetup{pdftitle={\thetitle}}%
\hypersetup{pdfsubject={\thesubject}}%
\hypersetup{pdfkeywords={\thekeywords}}%
}

\numberwithin{equation}{section}

\makeatletter\newcommand*{\etal}{%
    \@ifnextchar{.}%
        {et\penalty50\ al}%
        {et\penalty50\ al.\@\xspace}%
}\makeatother
\usepackage{xspace}

\makeatletter\newcommand*{\etc}{%
    \@ifnextchar{.}%
        {etc}%
        {etc.\@\xspace}%
}\makeatother

\newcommand{\nn}{\nonumber}

\newcommand{\be}{\begin{equation}}
\newcommand{\ee}{\end{equation}}
\newcommand{\bea}{\begin{eqnarray}}
\newcommand{\eea}{\end{eqnarray}}

\newcommand{\brk}[1]{(#1)}

\newcommand{\bigbrk}[1]{\bigl(#1\bigr)}
\newcommand{\Bigbrk}[1]{\Bigl(#1\Bigr)}

\newcommand{\sbrk}[1]{[#1]}

\newcommand{\bigsbrk}[1]{\bigl[#1\bigr]}

\newcommand{\Bigsbrk}[1]{\Bigl[#1\Bigr]}

\newcommand{\brc}[1]{\{#1\}}

\newcommand{\bigbrc}[1]{\bigl\{#1\bigr\}}
\newcommand{\Bigbrc}[1]{\Bigl\{#1\Bigr\}}

\newcommand{\abs}[1]{|#1|}
\newcommand{\bigabs}[1]{\bigl|#1\bigr|}

\newcommand{\supbrk}[1]{^{\brk{#1}}}
\newcommand{\cc}[2]{c\supbrk{#1}_{#2}}
\newcommand{\cct}[2]{\tilde{c}\supbrk{#1}_{#2}}

\newcommand{\eps}{\varepsilon}

\newcommand{\superN}{\mathcal{N}}

\newcommand{\dlog}{d \operatorname{log}}
\newcommand{\Li}{\operatorname{Li}}

\newcommand{\Z}{\mathbb{Z}}

\newcommand{\order}[1]{\mathcal{O}(#1)}

\newcommand{\subrm}[1]{_{\text{#1}}}
\newcommand{\dd}{\mathrm{d}}

\newcommand{\gcusp}{\Gamma\subrm{cusp}}

\newcommand{\EulerGamma}{\gamma_{\mathrm{E}}}
\newcommand{\id}{\operatorname{id}}
\newcommand{\Wmhv}{\mathcal{W}}
\newcommand{\Wnmhv}{W}
\newcommand{\Integers}{\mathbb{Z}}
\newcommand{\Rationals}{\mathbb{Q}}

\newcommand{\Zetas}{\mathcal{Z}}

\newcommand{\defas}{:=}

\newcommand{\stf}{f}
\newcommand{\estf}{\check{f}}
\newcommand{\ym}{1 - y}
\newcommand{\yinv}{\check{y}}
\newcommand{\op}{\mathcal{O}}
\newcommand{\logx}{\log\brk{x}}
\newcommand{\blogx}{\bigbrk{\logx}}

\newcommand{\blogym}{\bigbrk{\log\brk{\ym}}}
\newcommand{\cpr}{\Delta}
\newcommand{\diagSt}{\mathcal{D}}
\newcommand{\ndiagSt}{\mathcal{N}}
\newcommand{\up}[1]{u^+_{#1}}
\newcommand{\um}[1]{u^-_{#1}}
\newcommand{\Emhv}{\mathcal{E}}
\newcommand{\Enmhv}{\mathrm{E}}
\newcommand{\Efun}{E}
\newcommand{\Etfun}{\tilde{E}}
\newcommand{\Gfun}{\operatorname{\Gamma}}
\newcommand{\im}{\mathrm{i}}
\newcommand{\funspace}{\mathcal{H}^{\mathrm{DS}}}
\newcommand{\polyP}{\mathcal{P}}
\newcommand{\polyQ}{\mathcal{Q}}
\newcommand{\polyR}{\mathcal{R}}

\newcommand{\alphabet}{\mathcal{A}}
\newcommand{\intTensor}[1]{\mathcal{I}_{#1}}
\newcommand{\intSet}{\mathcal{S}}

\newcommand{\expX}{\mathcal{X}}
\newcommand{\GL}{\operatorname{GL}}
\newcommand{\kernel}{\operatorname{Ker}}
\newcommand{\weight}{p}
\newcommand{\intpath}{\gamma}

\newcommand{\cupdot}{\mathbin{\dot{\cup}}}

\newcommand{\Nopt}{N_{\text{opt}}}
\newcommand{\Gx}{\mathcal{G}}
\newcommand{\emptyword}{\epsilon}

\newcommand{\cA}{\begin{cal}A\end{cal}}

\newcommand{\cN}{\begin{cal}N\end{cal}}
\newcommand{\cO}{\begin{cal}O\end{cal}}

\makeatletter
\newcommand*\bigcdot{\mathpalette\bigcdot@{1}}
\newcommand*\smtimes{\mathpalette\smtimes@{.7}}
\newcommand*\bigcdot@[2]{\mathbin{\vcenter{\hbox{\scalebox{#2}{$\m@th#1\bullet$}}}}}
\newcommand*\smtimes@[2]{\mathbin{\vcenter{\hbox{\scalebox{#2}{$\m@th#1\times$}}}}}
\makeatother

\def\Li{{\rm Li}}
\def\cN{{\mathcal N}}
\def\cO{{\mathcal O}}

\def\to{\rightarrow}

\newcommand{\soft}[1]{\textsc{#1}}
\newcommand{\mathematica}{\soft{Mathematica}}

\newcommand{\filename}[1]{\texttt{#1}}
\newcommand{\code}[1]{\texttt{#1}}
\newcommand{\ancillary}[1]{\texttt{#1}}

\begin{document}


\title{Hexagon Bootstrap in the Double Scaling Limit}

\myabstract{We study the six-particle amplitude in planar $\mathcal{N} = 4$ super Yang-Mills theory in the double scaling (DS) limit, the only
nontrivial codimension-one boundary of its positive kinematic region. We construct the relevant function space, which is significantly constrained due to the extended Steinmann relations, up to weight 13 in coproduct form, and up to weight 12 as an explicit polylogarithmic representation. Expanding the
latter in the collinear boundary of the DS limit, and using the Pentagon Operator Product Expansion, we compute the non-divergent coefficient of a certain component of the
Next-to-Maximally-Helicity-Violating  amplitude through weight 12 and eight loops. We also specialize our results to the overlapping origin limit, observing a general pattern for its
leading divergences.}

\subject{Mathematical Physics, Scattering Amplitudes}



\ifarxiv
\noindent
\mbox{}\hfill DESY 20-221
\fi

\author{%
Vsevolod Chestnov,\texorpdfstring{${}^a$}{}
Georgios Papathanasiou\texorpdfstring{${}^b$}{}
}

\hypersetup{pdfauthor={\theauthor}}

\vfill

\begin{center}
{\Large\textbf{\mathversion{bold}\thetitle}\par}

\bigskip

\textsc{\theauthor}

\medskip

{\itshape
$^{a}$II. Institut f\"ur Theoretische Physik, Universit\"at Hamburg,\\
Luruper Chaussee 149, 22761 Hamburg, Germany

\medskip

$^{b}$DESY Theory Group, DESY Hamburg,\\
Notkestra\ss e 85, D-22603 Hamburg, Germany
}

\medskip

{\ttfamily\brc{%
\href{mailto:vsevolod.chestnov@desy.de}{vsevolod.chestnov},%
\href{mailto:georgios.papathanasiou@desy.de}{georgios.papathanasiou}%
}\href{vsevolod.chestnov@desy.de,georgios.papathanasiou@desy.de}{@desy.de}}
\par\vspace{1cm}

\textbf{Abstract}\vspace{5mm}

\begin{minipage}{12.4cm}
\theabstract
\end{minipage}

\vfill

\end{center}

\fillpdfdata

\newpage

\providecommand{\microtypesetup}[1]{}
\microtypesetup{protrusion=false}
\setcounter{tocdepth}{2}
\pdfbookmark[1]{\contentsname}{contents}
\tableofcontents
\microtypesetup{protrusion=true}



\section{Introduction}
\label{sec:intro}

The exact description of quantum interactions is one of the outstanding questions in theoretical physics. Our best hope for answering this question is
in the realm of the simplest interacting quantum field theory in four dimensions, $\cN=4$ super Yang-Mills theory
(SYM)~\cite{Brink:1976bc,Gliozzi:1976qd}.  Indeed, in the large-color or planar limit~\cite{tHooft:1973alw} the integrability of the theory has
enabled the determination of the scaling dimension spectrum of all its single-trace operators beyond perturbation theory, see for example the
reviews~\cite{Beisert:2010jr,Levkovich-Maslyuk:2019awk}.

For the quantities actually encoding the outcome of particle interactions, namely the scattering amplitudes, which in this theory also happen to be
equivalent to null polygonal Wilson loops~\cite{Alday:2007hr,Drummond:2007aua,Brandhuber:2007yx}, integrable structures are currently known to emerge
only in certain corners of the space of kinematics. The best-understood such corner is a certain collinear limit~\cite{Alday:2010ku}, whereby every
term in the series expansion of the Wilson loop or amplitude with respect to the kinematic variables that become small in this limit, can be mapped to
an exactly solvable flux tube~\cite{Basso:2013vsa, Basso:2013aha, Basso:2014koa, Basso:2014nra, Belitsky:2014sla, Belitsky:2014lta, Basso:2014hfa,
Basso:2015rta, Basso:2015uxa, Belitsky:2016vyq}. The flux tube description, known as the Wilson loop (or Pentagon) Operator Product Expansion (OPE),
is complete for the first nontrivial amplitude of the theory, which as a consequence of dual conformal symmetry (see \cite{Drummond:2010km} for a
review) has multiplicity $n=6$\footnote{At higher multiplicity, one final building block known as the ``matrix part'' is still missing.}.

In order to obtain closed expressions for this six-particle or `hexagon' amplitude, which will be the focus of this article, both at finite coupling
and in general kinematics, one would thus have to resum the aforementioned kinematic expansion. A strategy to achieve this ambitious goal, would be to
divide it into two simpler steps: First resum the kinematics order by order in perturbation theory with respect to the planar coupling $g$, and then
resum the perturbative series\footnote{Alternatively, one could start from the strong- instead of the weak-coupling regime,
see~\cite{Fioravanti:2015dma,Bonini:2015lfr,Bonini:2016knr,Bonini:2017gwt,Bonini:2018mkg} for work in this direction.}. Indeed, a great deal is known
about the class of polylogarithmic functions the first step evaluates to, thus greatly facilitating its realization. And it is not unreasonable to
expect that these polylogarithms can be in turn resummed to more complicated functions of hypergeometric type, as has been the case with certain
integrals contributing to the amplitude~\cite{Caron-Huot:2018dsv}, thanks to the existence of differential equations relating different perturbative
orders.

The task of resumming the perturbative OPE series was initiated in \cite{Drummond:2015jea}\footnote{For more recent work on the weak-coupling OPE resummation, see also \cite{Cordova:2016woh,Lam:2016rel,Bork:2019aud,Bork:2020aut}.}, also building on the earlier work
\cite{Papathanasiou:2013uoa,Papathanasiou:2014yva}, under one additional simplification: Starting from the one-dimensional collinear limit, the
kinematics was resummed to the ``double-scaling'' (DS) limit \cite{Gaiotto:2011dt,Basso:2014nra}, instead of the full three-dimensional space of
general kinematics of the hexagon. The double-scaling limit is distinguished by the fact that it is the only codimension-one boundary of the region of
positive kinematics, where the (appropriately normalized \cite{Bern:2005iz}) amplitude is nonvanishing, modulo its discrete symmetries. The positive
region, first considered in the context of amplitude integrands~\cite{Arkani-Hamed:2016byb} and then adapted to the space of external kinematics
in~\cite{Golden:2013xva}, is part of the Euclidean region, where amplitudes are free of branch points. Indeed, as is reviewed in
e.g.~\cite{Drummond:2018dfd}, the only other codimension-one boundary is the soft or equivalently multi-Regge limit, where the amplitude is
nonvanishing only after analytically continuing away from the Euclidean region.

From the point of view of the flux tube description, the double scaling limit is advantageous because only a simpler subset, of so-called
same-helicity gluon excitations contribute. These are not charged under the internal symmetries of the theory, and are labeled by a particle number
$N$, which also corresponds to the dimensionality of their all-loop integral representation. In~\cite{Drummond:2015jea}, in was in particular the
$N=1$ excitations which were considered, and it was realized that existing nested summation algorithms~\cite{Moch:2001zr} allow their explicit
evaluation in terms of two-dimensional harmonic polylogarithms~\cite{Gehrmann:2000zt}, or more precisely their subset associated to the $A_2$ cluster
algebra~\cite{Golden:2013xva}, in principle at any loop order.

In this work, we take the next step and study the $N=2$ gluon OPE excitations. These give rise to significantly more complicated integrals where, to
the best of our knowledge, no direct method for their evaluation is available to date (see \cite{article} for the current state of the art with
respect to nested summation technology), despite the fact that they are expected to lie in the same space of functions as the $N=1$ excitations.
Instead, we will rely on the bootstrap philosophy, where one first constructs the expected space of functions, and then locates the physical quantity
in question within this space. This approach has been first applied in the similar setting of the multi-Regge limit in~\cite{Dixon:2012yy}, and more recently it has been very successful for determining perturbative six- and seven-particle amplitudes  in planar
$\cN=4$ SYM in general
kinematics~\cite{Dixon:2011pw,Dixon:2011nj,Dixon:2013eka,Dixon:2014voa,Dixon:2014iba,Drummond:2014ffa,Dixon:2015iva,Caron-Huot:2016owq,Dixon:2016nkn,Drummond:2018caf,Caron-Huot:2019vjl,Caron-Huot:2019bsq,Dixon:2020cnr}.
In particular, the former are known through six and seven loops in the Next-to-Maximally Helicity-Violating (NMHV) and Maximally-Helicity-Violating
(MHV) configuration, respectively, whereas the latter are known through four loops, see also the recent review~\cite{Caron-Huot:2020bkp}.

First, we will thus develop the hexagon bootstrap in the simplified setting of the double scaling limit. We will see that while the limit breaks some
of the symmetries of the amplitude, it still preserves important analytic properties that tame the growth of the space of relevant functions with
respect to the weight, namely the number of iterated integrations defining them. Among these properties, a special role will be played by the extended
Steinmann relations~\cite{Caron-Huot:2018dsv,Caron-Huot:2019bsq}, which generalize the ordinary Steinmann
relations~\cite{Steinmann,Steinmann2,Cahill:1973qp} so as to forbid not only double, but also multiple discontinuities in overlapping channels. While
these channels are normally associated to Mandelstam invariants, for functions with physical branch cuts the extended Steinmann relations imply the
absence of discontinuities also with respect to more general kinematic variables, as also predicted by the principle of cluster
adjacency~\cite{Drummond:2017ssj,Drummond:2018dfd}. Specifically in the double-scaling limit, we find that two of such generalized discontinuities are
forbidden. With their aid, we will be able to construct the corresponding space of `Extended Steinmann Double-Scaling' (DS for short) functions to
weight 12 explicitly, and to weight 13 when the functions are specified iteratively in terms of their first derivatives (or coproducts, see
\cite{Duhr:2014woa} for a review).

Then, we will proceed to uniquely identify the contribution of $N=1,2$ gluon OPE excitations inside the DS space. On the one hand, we will work out
the expansion of our functions in the collinear limit, and on the other hand we will compare that to the sum representation of the OPE predictions,
obtained from their original integral form with the help of Cauchy's residue theorem, and organized into finite coefficients multiplying divergent
logarithms in the limit. So as to be able to provide useful boundary data and checks to the amplitude bootstrap in general kinematics, here we will be
focusing on the contributions to the NMHV (super)amplitude,  which carries both rational and transcendental dependence on the kinematics. Given that
the $N>3$ OPE contrinutions only start contributing at higher loop orders, in this manner we will be able to determine the finite coefficients in the
DS limit of the NMHV hexagon for a particular, so-called (1111) component of its rational depenendence through weight 12 and eight loops.

With these results at hand, it is possible to study further interesting subspaces of the DS limit. As an example, we will indeed also specialize them
to a DS boundary point known as the origin limit~\cite{Caron-Huot:2019vjl}, where a similar OPE resummation strategy as the one employed here, has led
to finite-coupling conjectures for the form of the MHV amplitude~\cite{Basso:2020xts}, which exhibits a Sudakov-like exponentiation. As was observed
in the latter paper, and we confirm here, the NMHV amplitude no longer exhibits this exponentiation. Nevertheless, we observe a general pattern for
its leading divergence at the origin, which may be valid to all loops.

Finally, we briefly address the question of how `minimal' the DS function space is, namely whether it also contains redundant functions which are
not present in the amplitude or its derivatives. By comparing with the space spanned by the latter, we notice that non-constant redundant functions
already start appearing at weight three. Understanding the reason for this redundancy, and further refining our space by eliminating it, are
interesting questions that we leave for the future. Perhaps more importantly, in follow-up work we look forward to using the host of explicit results
we have obtained, in order to develop new direct evaluation algorithmms for the two-gluon OPE contributions, which may also be more broadly applicable
to perturbative quantum field theory.  The presence of a similar successful paradigm, where the knowledge of certain double pentagon ladder
integrals~\cite{ArkaniHamed:2010kv} to high loop order~\cite{Caron-Huot:2018dsv} subsequently led to new computational methods~\cite{McLeod:2020dxg},
is very encouraging in this respect.

This paper is organized as follows. In~\secref{sec:review}, we start by reviewing the essential analytic properties of the six-particle amplitude
in general kinematics, and then move on to work out their implications in the DS limit. Relying on these properties, in~\secref{sec:bootstrap} we
describe the construction of our DS function space and its expansion in the collinear limit. The latter is to be compared with the predictions of the
Wilson loop OPE, discussed in~\secref{sec:hex}. The thus obtained new results on the on the NMHV amplitude in the DS limit are presented
in~\secref{sec:results}. Finally, we have included two appendices with further details on our DS space construction. Our results are also attached
as computer-readable files accessible at \cite{CPrepos}.

\section{The six-gluon amplitude in the double-scaling limit}
\label{sec:review}

In this section we deduce the analytic properties of the six-particle amplitude in the DS limit. In subsection \ref{ssec:notation} we first review
some background information on the normalization of the amplitude, its analytic structure in general kinematics, and the class of multiple
polylogarithms encoding it. Then in subsection \ref{ssec:DS_general} we define the DS limit, and describe the potential amplitude singularities in the
limit. Subsection \ref{ssec:IntegrStein} continues with the analysis of the property of integrability and of the extended Steinmann relations,
especially illustrating the constraining power of the latter. Finally in subsection \ref{ssec:branchCutCond} we derive additional conditions obeyed by
the amplitude and its derivatives in the DS limit, stemming from the absence of unphysical branch cuts.

\subsection{Analytic structure of the normalized amplitude}
\label{ssec:notation}

The infrared divergence structure of the six-particle amplitude is well-understood~\cite{Bern:2005iz} and can be factored out, giving rise to a finite normalized amplitude. Different conventions on this normalization reflect the freedom to also absorb finite terms in this infrared-divergent factor, which here we will choose as the BDS-like ansatz $A^{\text{BDS-like}}_{6}$~\cite{Alday:2009dv}. Its precise form will not be important for our purposes, and instead we will be focusing on the BDS-like normalized amplitudes of the two inequivalent helicity configurations,
\be\label{eq:cEEDef}
\Emhv \defas \frac{A_{6,\textrm{MHV}}}{\,A^{(0)}_{6,\textrm{MHV}}\,A^{\text{BDS-like}}_{6}}\,,\quad
\Enmhv \defas \frac{A_{6,\textrm{NMHV}}}{\,A^{(0)}_{6,\textrm{MHV}}\,A^{\text{BDS-like}}_{6}}\,,
\ee
where we have also divided out by their known tree-level contribution. While $\Emhv$ is given just by a single bosonic function of the kinematics, $\Enmhv$ may in turn be further decomposed as
\begin{align}
    \Enmhv = \tfrac12 \Bigsbrk{
        \bigbrk{\brk{1} + \brk{4}} \Efun\brk{u, v, w}
        + \bigbrk{\brk{2} + \brk{5}} \Efun\brk{v, w, u}
        + \bigbrk{\brk{3} + \brk{6}} \Efun\brk{w, u, v}
        \nn\\
        + \bigbrk{\brk{1} - \brk{4}} \Etfun\brk{{u}, {v}, {w}}
        + \bigbrk{\brk{2} - \brk{5}} \Etfun\brk{{v}, {w}, {u}}
        + \bigbrk{\brk{3} - \brk{6}} \Etfun\brk{{w}, {u}, {v}}
    }.
    \label{eq:EnmhvDef}
\end{align}
where $(1) \defas [23456]$ and its cyclic permutations denote the so-called $R$-invariants~\cite{Drummond:2008vq,Drummond:2008bq,ArkaniHamed:2010kv},
namely rational terms of the kinematics and the Grassmann variables, encoding the superconformal and dual superconformal symmetry of the amplitude.

The two functions $\Efun$ and $\Etfun$ introduced in eq. \eqref{eq:EnmhvDef}, as well as the entire MHV amplitude $\Emhv$, are bosonic functions of
the kinematical data, which can be conveniently parametrised by a set of three cross-ratios:
\be
    \brc{u, v, w} \defas \Bigbrc{
        \frac{s_{12} s_{45}}{s_{123} s_{345}},
        \frac{s_{23} s_{56}}{s_{234} s_{123}},
        \frac{s_{34} s_{61}}{s_{345} s_{234}}
    },
    \label{eq:uvwDef}
\ee
which are invariant under the parity transformation. Evidence from all explicit results to date, as well as from the analysis of the integrand~\cite{ArkaniHamed:2012nw} (note however the caveats pointed out in \cite{Brown:2020rda}) that the order $g^{2 L}$ ($L$-loop) contribution in the weak coupling expansion of $\Emhv, \Efun, \Etfun$ can be expressed in terms of \emph{multiple polylogarithms} (MPLs) \cite{Chen:1971, Chen:1977oja, G91b,Goncharov:1998kja} \brk{see also the review~\cite{Duhr:2014woa}}
of transcendental weight $\weight = 2L$. A function $F\supbrk{\weight}$ is said to be an MPL of weight $\weight$ if its total differential obeys
\be
    dF\supbrk{\weight} =
    \sum_{\beta\in{\alphabet}} \bigbrk{F\supbrk{\weight - 1}}^{\beta} \> \dlog\brk{\beta}\,,
    \label{dFPhi}
\ee
such that that $\bigbrk{F\supbrk{\weight - 1}}^{\alpha}$ is an MPL of weight $\brk{\weight - 1}$ and so on, where the recursive definition terminates with the usual logarithms ($\weight=1$) on the left-hand side, and rational numbers ($\weight=0$)
as coefficients of the total differentials on the right-hand side. The set $\alphabet$ of arguments of dlog forms is called the \emph{symbol alphabet},
and it encodes positions of possible branch points of the $F\supbrk{\weight}$ function.

This recursive nature of the differential of MPLs is a part of a deeper structure revealed by the \emph{coproduct} (more precisely, \emph{coaction}) $\Delta$
\cite{Goncharov:2001iea,Goncharov:2005sla,Brown:2011ik,Duhr:2012fh}, which, very roughly, decomposes an MPL of weight $\weight$ into a sum of tensor products of MPLs of lower
weight.
In particular, the total differential \eqref{dFPhi} is essentially equivalent to the $\brc{\weight - 1, 1}$ component of $\Delta$,
\be
    \Delta_{\weight - 1,1} F\supbrk{\weight} =
    \sum_{\beta \in \alphabet} \bigbrk{F\supbrk{\weight-1}}^{\beta} \otimes
    \bigsbrk{\log\brk{\beta} \mod\,(\im \pi)}\,.
    \label{eq:Deltan1}
\ee
Further considering the total differential of $F\supbrk{\weight-1}$, or equivalently the analogue of eq. \eqref{eq:Deltan1} for the latter, then yields the $\{\weight-2,1,1\}$ component of the coproduct,
\begin{align}
    \Delta_{\weight-2,1,1}F\supbrk{\weight} =
    \sum_{\alpha,\beta\in{\alphabet}} \bigbrk{F\supbrk{\weight-2}}^{\alpha,\beta} \otimes \log \alpha \otimes \log \beta\,,
    \label{Delta11}
\end{align}
where we will also refer to the leftmost factor on the right-hand side as the `double coproduct'. In the above formula, and in what follows, identification of logarithmic factors up to $\im \pi$ is implied in all but the first slot of the tensor product. We can also continue the decomposition of the leftmost coproduct factor, until we reach the maximal, $p$ times iterated coproduct $\{1,\ldots,1\}$, which is also known as the \emph{symbol}.

Another very usefull point of view on the MPLs stems from their integral representation: an MPL is defined to be a $\Rationals$-linear
combination of the following iterated integrals \brk{sometimes refered to as ``hyperlogarithms''}:
\begin{align}
    G\brk{a_1, \ldots, a_\weight; z} \defas \int_0^z G\brk{a_2, \ldots, a_\weight; t} \; \frac{\dd t}{t - a_1}, \quad
    G\brk{; z}  \defas 1,
\end{align}
where the special case of only zero arguments is covered by this rule:
\begin{align}
    G\brk{\underbrace{0, \ldots, 0}_{\weight}; z} \defas \frac{\bigbrk{\log{z}}^\weight}{\weight!}.
    \label{eq:GdefLog}
\end{align}

The space of so-called \emph{hexagon functions} containing $\brc{\Emhv, \Efun, \Etfun}$ and their coproducts is in fact a much smaller subspace of all MPLs, due to additional physical and mathematical constraints that the amplitude satisfies. The idea of the hexagon function bootstrap~\cite{Dixon:2011pw,Dixon:2011nj,Dixon:2013eka,Dixon:2014voa,Dixon:2014iba,Dixon:2015iva,Caron-Huot:2016owq,Caron-Huot:2019vjl,Caron-Huot:2019bsq} is to first construct this space, from its basis then form an ansatz for the amplitude, and finally find a unique solution for the latter by comparing it to the behavior of the amplitude in various kinematic limits, known by other means. Below we will briefly review the additional analytic properties of hexagon functions, which we will then specialize to the double-scaling limit.

\medskip

\noindent
\emph{Symbol alphabet}. The space of hexagon functions containing $\brc{\Emhv, \Efun, \Etfun}$ are MPLs as defined in eq.~\eqref{dFPhi}, whose letters are drawn from the  following list~\cite{Dixon:2011pw}:
\be
    \cA^\text{hex}=\{u,v,w,1-u,1-v,1-w,y_u,y_v,y_w\}\,.
    \label{eq:hex_alphabet}
\ee
Apart from the parity-even cross ratios of eq.~\eqref{eq:uvwDef}, we have also introduced parity odd letters that are expressed in terms of
the latter \brk{$u_1=u,u_2=v,u_3=w$ and similarly for $y_i$} as
\be
    y_i \defas \frac{u_i-z_+}{u_i-z_-}\,,\quad
    z_\pm \defas \frac{1}{2}(-1+u+v+w\pm \sqrt{\delta})\,,\quad
    \delta \defas (1-u-v-w)^2-4uvw\,.
    \label{eq:yDef}
\ee
Zeros of these expressions label the
possible locations of the branch cut singularities of MPLs in the ansatz.

\medskip

\noindent
\emph{First entry condition}. Locality dictates that in the Euclidean region, amplitudes can only develop singularities at its boundary, where any of the cross ratios tend to zero. This implies that only the first three letters in eq. \eqref{eq:hex_alphabet} are allowed to appear in the first entry of the symbol~\cite{Gaiotto:2011dt}, or equivalently that the weight-one space of hexagon functions consists of
\be\label{eq:FirstEntry}
F^{(1)}\in\brc{\log u, \log v, \log w}\,.
\ee

\medskip

\noindent
\emph{Integrability conditions}. Any well-defined hexagon function $F$ must satisfy
\be
    \frac{\partial^2 F}{\partial x_i\partial x_j}\ =\
    \frac{\partial^2 F}{\partial x_j\partial x_i} \,, \qquad i \neq j,
    \label{eq:commdoublederiv}
\ee
for any choice of variables $x_1, x_2,x_3$ parametrizing the kinematics, such as the cross ratios \eqref{eq:uvwDef}. Given the relation between total differentials and coproduct components \eqref{dFPhi,eq:Deltan1}, these translate to linear relations for the left factors of the double coproduct of eq. \eqref{Delta11}, known as the integrability conditions.

\medskip

\noindent
\emph{Extended Steinmann relations}. Basic principles of quantum field theory prohibit virtual particles in any physical process to simultaneously become on-shell in two overlapping channels~\cite{Steinmann,Steinmann2,Cahill:1973qp}. This can be translated into the vanishing of certain double discontinuities of the associated physical quantity, or equivalently into restrictions on the first two entries of its symbol (if it is described by MPLs). In~\cite{Caron-Huot:2018dsv,Caron-Huot:2019bsq} it was realized that these relations in fact hold not on just the first two, but on any adjacent letters in the symbol. These `extended Steinmann relations' may be equivalently be stated in terms of the double coproducts of any hexagon function $F$ as
\begin{align}
    F^{v, u} + F^{v, w} + F^{w, u} + F^{w, w} = 0,
    \label{eq:extendedSt}
\end{align}
together with two other cyclic permutations $u \to v \to w \to u$. When combined with the other analytic properties mentioned thus far, they also automatically imply the following double coproduct relations,
\begin{align}
F^{{1 - u,v}}+F^{{1 - u,w}}+F^{{1 - u,1 - v}}+F^{{1 - u,1 - w}}&=0\\
F^{{w,y_w}}+F^{{1 - v,y_w}}+F^{{1 - w,y_w}}-F^{{v,y_v}}-F^{{1 - v,y_v}}-F^{{1 - w,y_v}}+F^{{v,y_w}}-F^{{w,y_v}}&=0\\
F^{{1 - u,y_v}}-F^{{1 - u,y_u}}+F^{{y_v,1 - w}}-F^{{y_w,1 - w}}&=0\\
F^{{1 - v,1 - u}}+F^{{y_u,y_v}}+F^{{y_w,y_w}}-F^{{y_u,y_w}}-F^{{y_w,y_v}}&=0\label{eq:extraDoubleCop}
\end{align}
plus cyclic permutations. The above relations, eqs. \eqref{eq:extendedSt}-\eqref{eq:extraDoubleCop}, on integrable functions are also predicted by the principle of cluster
adjacency~\cite{Drummond:2017ssj,Drummond:2018dfd}, which relies on the fact that the letters \eqref{eq:hex_alphabet} are isomorphic to the variables of a mathematical object known as $A_3$ cluster algebra~\cite{Golden:2013xva}. Part of the structure of this object is the arrangement of the cluster variables in overlapping sets, which may be naturally translated to double coproduct relations.

\medskip

\noindent
\emph{Branch cut conditions and transcendental constants}. All of the above double coproduct relations in fact define hexagon functions only up to constants times weight-one logarithms. Not every value of these constants is allowed however, since it may lead to beyond-the-symbol terms with unphysical branch cuts. As was discussed in \cite{Dixon:2013eka,Dixon:2015iva}, one way to eliminate this possibility is to additionally impose that hexagon functions are well-behaved in certain kinematic limits.

Here we will adopt the choice of the aforementioned papers, and consider the soft limit
\be
    u,w\to 0,v\to 1\,\,\, \text{with}\,\,\, \frac{u}{1-v}, \frac{w}{1-v}\,\,\, \text{fixed}\,\,\,\Leftrightarrow\,\,\, y_v\to 1  \,\,\,\text{with}\,\,\, y_u,y_w \,\,\, \text{fixed}\,,
    \label{eq:softlimit1}
\ee
where the $\brc{p - 1, 1}$ coproduct component of hexagon functions  should obey
\be
    F^{1-v}\big|_{y_v\to 1}=F^{y_u}\big|_{y_v\to 1}=F^{y_w}\big|_{y_v\to 1}=0\,.
    \label{eq:hexBranchCuts}
\ee
These conditions need to also be supplemented with the set of transcendental constants that we include as independent functions in our basis, and then they determine the precise rational coefficients these multiply weight-1 logarithms with. As is also reviewed in~\cite{Caron-Huot:2020bkp}, on general grounds these constants should be drawn from multiple zeta values (MZV), and in~\cite{Caron-Huot:2019vjl,Caron-Huot:2019bsq} it was furthermore conjectured that only their subset of even ordinary
zeta values of weight at least four,
\begin{align}
    \brc{\zeta_4, \zeta_6, \zeta_8, \zeta_{10}, \zeta_{12}, \ldots}
    \label{eq:cosmicMZV}
\end{align}
is necessary for the six-particle amplitude and its derivatives at any loop order.

Let us now proceed to work out the consequences of these analytic properties of the amplitude in the double-scaling limit, which will be the focus of this article.

\subsection{The double-scaling (DS) limit}
\label{ssec:DS_general}

\begin{figure}[t]
    \centering
    \includegraphics[align=c]{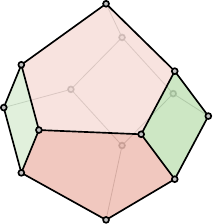}
    \caption{
        The $A_3$ Stasheff associahedron, relevant for six-particle scattering. The positive region in the space of kinematics amounts to the interior
        of the polytope, whereas pentagonal faces correspond to double-scaling limits, and square faces to soft (or equivalently multi-Regge) limits.
        The two {\color{red1} red} pentagonal faces represent the two possible double scaling limits DS1 and DS2 that are considered in this work.
    }
    \label{fig:A3Stasheff}
\end{figure}

The double-scaling limit~\cite{Gaiotto:2011dt,Basso:2014nra,Drummond:2015jea} of the six-particle amplitude in the planar $\superN = 4$ SYM is generally defined as the limit where one of the three cross ratios~\eqref{eq:uvwDef} goes to zero, whereas the remaining two are held fixed. In this work we will choose the particular orientation
\be
\text{DS limit:}\quad w\to 0\,\,\,\text{with}\,\,\,u,v\,\,\, \text{fixed},
\ee
where it is easy to show that the discriminant of eq. \eqref{eq:yDef} reduces to
\be
\sqrt{\delta}
\xrightarrow[\text{DS}]{}
\pm (1-u-v)\left[1-\frac{(1-u-v+2 u v) w}{(1-u-v)^2}\right]+\cO (w^2)\,,
\label{eq:deltaLim}
\ee
with different signs denoting the two possible choices of the square root branch, such that the odd letters become
\be\label{eq:y_DSlimit}
y_u\to\left[\frac{u}{1-v}\right]^{\pm 1}\,,\quad y_v\to\left[\frac{v}{1-u}\right]^{\pm 1}\,,\quad y_w\to\left[\frac{(1-u)(1-v)w}{(1-u-v)^2}\right]^{\pm 1}\,.\quad
\ee
Comparing with eq. \eqref{eq:hex_alphabet}, we thus see that in this limit hexagon functions $F$ reduce to divergent logarithms in $w$, times `DS functions' $\stf$ with alphabet

\be\label{eq:uv_alphabet}
\{u,v,1-u,1-v,1-u-v\}\,.
\ee
By virtue of eq. \eqref{eq:y_DSlimit}, the $\{\weight-1,1\}$ coproducts of the DS functions are related to the hexagon functions $F$ as follows:
\be\label{eq:hextoDSCoproducts}
\begin{aligned}
\stf^u&=F^u\pm F^{y_u}\,,\\
\stf^v&=F^v\pm F^{y_v}\,,\\
\stf^{1-u}&=F^{1-u}\pm (F^{y_w}- F^{y_v})\,,\\
\stf^{1-v}&=F^{1-v}\pm (F^{y_w}- F^{y_u})\,,\\
\stf^{1-u-v}&=\mp 2 F^{y_w}\,.\end{aligned}
\ee
(plus $\stf^{\prime w}=F^w\pm F^{y_w}$ if $\stf'$ denotes the products of $\stf$ with $\log w$). As it was with eq. \eqref{eq:deltaLim}, the $\pm$
sign here is due to the fact that hexagon functions are in fact well-defined only in the double cover of the cross-ratio parametrization of the kinematics. In
other words, there really exist two $w\to 0$ double-scaling limits DS1 and DS2, that are related by a parity transformation, $y_i\to 1/y_i$. With the appropriate choice of independent kinematic variables, where the region of positive kinematics is the particular blowup of the unit cube shown in \figref{fig:A3Stasheff}, the two DS limits are the boundaries depicted in {\color{red1} red}. As is reviewed in e.g. \cite{Drummond:2018dfd}, the six-particle positive region has the topology of the Stasheff associahedron, closely related to the $A_3$ cluster algebra, whereas pentagonal faces represent its $A_2$ subalgebras.

In what follows, we will use an alphabet equivalent to \eqref{eq:uv_alphabet}, which arises naturally when evaluating the Wilson loop OPE
contributions to the DS limit . In particular, we define
\begin{align}
    x \defas -\frac{1-u-v}{v}\,,\quad
    y \defas \frac{1-v}{1-u-v}\,,
    \label{eq:xytou}
\end{align}
or conversely
\be\label{eq:utoxy}
u=\frac{x(1-y)}{1-xy}\,,\quad v=\frac{1}{1-xy}\,,
\ee
such the alphabet \eqref{eq:uv_alphabet} takes the equivalent form
\be
    \alphabet \defas \{x,y,1-x,1-y,1-x y\}\,,
    \label{eq:xy_alphabet}
\ee
which we will adopt from this point on. For completeness, we also quote how the $\brc{\weight - 1, 1}$ coproducts of the DS functions in the two equivalent alphabets are related by virtue of eq. \eqref{eq:xytou},
\begin{align}
    \stf^u&=\stf^{1-y}\,,\nn\\
    \stf^v&=-\stf^x-\stf^{1-x}-\stf^{1-xy}\,,\nn\\
    \label{eq:uvToxyCoprods}
    \stf^{1-u}&=\stf^{1-x}\,,\\
    \stf^{1-v}&=\stf^{y}\,,\nn\\
    \stf^{1-u-v}&=\stf^{x}-\stf^{y}-\stf^{1-y}\,.\nn
\end{align}

Finally, in the above choice of alphabet, it is evident that for DS functions, the first entry condition~\eqref{eq:FirstEntry} becomes
\be\label{eq:xy_FirstEntry}
f^{(1)}\in \brc{\log\brk{1-xy}, \log\brk{x(1-y)}}\,.
\ee

\subsection{Integrability and Extended Steinmann conditions}\label{ssec:IntegrStein}

After defining the DS limit, in the previous subsection we also analyzed the symbol letters and first entry condition of the space of functions expected to capture six-particle scattering in this limit. Here, we continue by deriving constraints on the double coproducts of these functions.

First, the integrability conditions are very simple to derive for the alphabet~\eqref{eq:xy_alphabet} we will be using from a single equation of the form \eqref{eq:commdoublederiv}, with $x_1=x$ and $x_2=y$. It gives rise to the following set of six equations,
\be\label{eq:integrability}
\begin{aligned}
    \stf^{x, y}-\stf^{y, x} =
    \stf^{1 - x, y}-\stf^{y, 1 - x} =
    \stf^{x, 1 - y}-\stf^{1 - y, x} = 0, \\
    \stf^{1 - x, 1 - x y}+\stf^{1 - x, 1 - y}-\stf^{1 - x y, 1 - x}-\stf^{1 - y, 1 - x} = 0, \\
    \stf^{1 - x, 1 - y}+\stf^{1 - x y, 1 - y}-\stf^{1 - y, 1 - x}-\stf^{1 - y, 1 - x y} = 0, \\
    \stf^{x, 1 - x y}-\stf^{1 - x, 1 - y}-\stf^{1 - x y, x}+\stf^{1 - x y, y}-\stf^{y, 1 - x y}+\stf^{1 - y, 1 - x} = 0.
\end{aligned}
\ee
Next, we examine the extended Steinmann relations, as well as the additional relations they imply on integrable functions with the alphabet~\eqref{eq:hex_alphabet} and first entries~\eqref{eq:FirstEntry}, eqs. \eqref{eq:extendedSt}-\eqref{eq:extraDoubleCop}. We find that only two linear combinations of these 15 equations survive in the DS limit, in particular
\begin{align}
    \stf^{1 - x, y} =
    \stf^{x, 1 - x y}+\stf^{1 - x, 1 - x y}+\stf^{1 - x y, 1 - x}+\stf^{1 - x y, 1 - x y}+\stf^{1 - x y, 1 - y}-\stf^{y, 1 - x y} = 0.
    \label{eq:steinmann}
\end{align}
Owing to their origin in general kinematics, we will denote eqs.~\eqref{eq:steinmann} as the `DS Extended Steinmann relations'.

One may be tempted to think that the DS Extended Steinmann relations may only marginally reduce the size of the relevant function space, due to their relatively small number. As we see in~\tabref{tab:symbols} however, their effect is in fact very significant: Focusing momentarily on the space of symbols, i.e. polylogarithmic functions modulo transcendental constants (these will be reinstated in the next subsection) with the alphabet \eqref{eq:xy_alphabet}, in the first line we quote their number as a function of their weight $p$, when the weight-one or first entry space is constrained as dictated by eq. \eqref{eq:xy_FirstEntry}. We notice that this number grows by roughly a factor of three at each weight (more precisely, it is equal to $3^{p - 1}+2^{p - 1}$).

On the second line of~\tabref{tab:symbols}, we display how many of the functions of the first line, additionally obey the DS Extended Steinmann relations~\eqref{eq:steinmann}. Evidently, their number now grows by roughly a factor of two instead of three. Thus the DS Extended Steinmann relations are responsible for a massive reduction in the size of the relevant function space, for example by more than 98\% for weight 13. They will thus be pivotal for constructing this function space to high weight,  as detailed in the next section, and for bootstrapping  new results for the six-particle NMHV amplitude in the DS limit, as presented in \secref{sec:results}.

\begin{table}
    \centering
\resizebox{\textwidth}{!}{\begin{tabular}{l|cc ccc ccc ccc ccc}
        \toprule
            weight $\weight$
             & 1 & 2 & 3 & 4 &  5 &  6 &  7 &  8 &  9 & 10 & 11 & 12 & 13 \\
        \midrule
First entry          & 2 & 5 & 13 & 35 & 97 & 275 & 793 & 2315 & 6817 & 20195 & 60073 & 179195 & 535537

\\
Ext. Stein.           & 2 & 4 & 9 & 19 & 39 & 78 & 154 & 302 & 591 & 1157 & 2269 & 4460 & 8788 \\
        \bottomrule
    \end{tabular}}
    \caption{The dimension of the space of symbols with physical branch cuts in the DS limit (first line), that additionally obey the extended Steinmann relations (second line).}
    \label{tab:symbols}
\end{table}

This concludes the analysis of all symbol-level constraints on the DS functions space. In the next subsection, we discuss the additional constraints necessary to promote them to functions.

\subsection{Branch cut conditions and transcendental constants}
\label{ssec:branchCutCond}
As briefly reviewed in subsections~\ref{ssec:notation} and~\ref{ssec:DS_general}, hexagon or DS functions are allowed to have branch points only when a cross ratio approaches zero (or infinity). Focusing on DS functions, one should thus require that they are free of branch point singularities when $u\to 1$, $ v\to 1$ or $u\to1-v$, or equivalently that their derivatives
are free of poles there. Given that
\be
\frac{\partial \stf}{\partial u}=\frac{\stf^u}{u}-\frac{\stf^{1-u}}{1-u}-\frac{\stf^{1-u-v}}{1-u-v}\,,
\ee
the physical branch cut conditions thus translate to
\be\label{eq:uv_branchcuts}
\stf^{1-u}\big|_{u\to 1}=\stf^{1-v}\big|_{v\to 1}=\stf^{1-u-v}\big|_{u\to 1-v}=0\,.
\ee

Let us now translate the above branch cut limits and conditions in our choice of independent variables~\eqref{eq:xytou} and symbol letters~\eqref{eq:xy_alphabet}. As far as the limits are concerned, we have
\begin{align}
    u&\to 1: x\to 1\,\,\text{with $y$ fixed}\,,\nn\\
    v&\to 1: y\to 0\,\,\text{with $x$ fixed}\,,\\
    u&\to 1-v: x\to 0\,\,\text{with $xy$ fixed}\,,\nn
\end{align}
whereas by virtue of~\eqref{eq:uvToxyCoprods} the branch cut conditions become
\be\label{eq:xy_branchcuts}
\stf^{y}\big|_{y\to 0}=\stf^{1-x}\big|_{x\to 1}=\stf^{x}-\stf^{y}-\stf^{1-y}\big|_{\substack{x\to 0\hfill\\xy\,\,\text{fixed}}}=0\,.
\ee

A potential subtlety with the branch cut conditions \eqref{eq:uv_branchcuts} or \eqref{eq:xy_branchcuts}, is that the limits they describe are not
just the edges of the pentagonal faces of the Stasheff polytope depicted in \figref{fig:A3Stasheff}, which are usually simpler to impose. This is
especially the case with the third limit/condition in \eqref{eq:xy_branchcuts}, which requires taking $y \to \infty$ simultaneously with $x \to 0$,
and thus lies outside of the $0 < x, y < 1$ square, where any function with the alphabet $\alphabet$ given in eq. \eqref{eq:xy_alphabet} is free of
branch cuts.

To avoid this complication, we will instead rely on the fact that the DS limit overlaps with the soft limit on the edge between a {\color{red1}
red} pentagonal and a {\color{green1} green} square face of the Stasheff polytope in \figref{fig:A3Stasheff}, where branch cut conditions for the
general hexagon function space have been derived.
From \eqref{eq:utoxy}, it is clear that the DS limit intersects with the soft limit \eqref{eq:softlimit1} for $x\to0$ with $y$ fixed.  Then, using
\eqref{eq:hextoDSCoproducts} and \eqref{eq:uvToxyCoprods} to relate the DS coproducts in the $\alphabet$ alphabet to
the hexagon function coproducts, and specializing to the double-scaling/soft overlap, we can show that by virtue of eq.~\eqref{eq:hexBranchCuts},
\be
\stf^x-\stf^{1-y}\big|_{x\to0}\,=F^{1-v}\mp (F^{y_w}+F^{y_u})\big|_{y_v\to 1}=0\,.
\ee
Instead of eq. \eqref{eq:xy_branchcuts}, we will thus choose to impose the following set of simpler branch cut conditions,
\begin{align}
    \stf^{y}\big|_{y \to 0}=
    \stf^{1-x}\big|_{x \to 1}=
    \stf^{x} - \stf^{1 - y}\big|_{x \to 0}=
    0\,.
    \label{eq:brcond}
\end{align}
Finally, let us come to the question of which transcendental constants we should include in our DS function space as independent basis elements. While only the subset \eqref{eq:cosmicMZV} of MZVs have been found to be necessary in general kinematics, it is easy to show that in the DS limit we will certainly also need $\zeta_2$. In particular, this is what the weight-two hexagon function $\text{Li}_2(1-1/w)=-G(0,1;1-1/w)$, see e.g.~\cite{Caron-Huot:2019bsq}, reduces to in the limit. A similar analysis at weight three indicates that we also need to include $\zeta_3$ as an independent constant. Given that these low-weight constants will be responsible for the bulk of non-trivial functions with vanishing symbols in our space, we will choose to be agnostic and include all MZVs, a basis of which is contained in appendix \ref{app:mzvs} through weight twelve, as independent functions in our space. We will then come back and examine the possible redundancy of our space in the closing  subsection~\ref{ssec:saturation}.

Summarizing, we define our space of Extended Steinmann DS functions, $\funspace$, to consist of MPLs with the alphabet of eq. \eqref{eq:xy_alphabet}, whose first entry space is restricted to~\eqref{eq:xy_FirstEntry}, and obeys the integrability~\eqref{eq:integrability} DS extended Steinmann~\eqref{eq:steinmann} and branch cut conditions~\eqref{eq:brcond}, also containing all MZVs as independent functions. In the next section, we describe the construction of this space in detail.
%

\section{Bootstrapping the DS Functions}
\label{sec:bootstrap}

The goal of this section is to construct a linear space of functions $\funspace$, which we refer to as the (Extended Steinmann) DS space, encoding
six-particle scattering in the DS limit, defined in the previous section. The DS space has a natural grading by transcendental weight,
\be
\funspace = \bigoplus_{\weight \ge 1} \funspace_\weight\,,
\ee where the $\funspace_\weight$ components
consist of linear combinations of polylogarithms and Multiple Zeta Values \brk{MZVs} of weight $\weight$. MZVs form a graded subalgebra of
their own,
\be \Zetas = \bigoplus_{\weight \ge 0} \Zetas_{\weight}\,,
\ee
with $\Zetas_0 = \Rationals$ and each component at weight $\weight$
has dimensionality $\abs{\Zetas_p} \defas \dim_{\Rationals} \Zetas_p$ with corresponding basis collected in the $\weight^{\text{th}}$ element of
eq. \eqref{eq:mzvbasis}.

In subsection~\ref{ssec:coprods}, we first explain how to build a coproduct representarion of $\funspace$, relying on the analytic properties worked
out in \secref{sec:review}. Then, in subsection~\ref{ssec:funs} we promote this representation to explicit expressions in terms of MPLs. Finally, in
subsection~\ref{ssec:expansions} carry out the $x \to 0$ series expansion of our functions, which will allow us to match them against predictions for
the amplitude in the collinear limit, as we will discuss in the next section.

\subsection{Solving the integrability and extended Steinmann constraints}
\label{ssec:coprods}
\begin{table}
    \centering
    \begin{tabular}{c|cccccc}
        \toprule
        weight              & $2$ & $3$ & $4$ & $5$ & $6$ & $7$ \\
        number              & $5$ & $12$ & $26$ & $56$ & $116$ & $236$ \\
        non-zero            & $12$ & $31$ & $96$ & $263$ & $901$ & $2.6 \times 10^3$ \\
        density             & $0.24$ & $0.10$ & $0.06$ & $0.04$ & $0.03$ & $0.019$ \\
        max                 & $1$ & $2$ & $2$ & $20$ & $40$ & $560$ \\
        \midrule
        weight              & $8$ & $9$ & $10$ & $11$ & $12$ & $13$ \\
        number              & $474$ & $943$ & $1867$ & $3686$ & $7270$ & $14295$ \\
        non-zero            & $9.9 \times 10^3$ & $39 \times 10^3$ & $.2 \times 10^6$ & $.8 \times 10^6$ & $3.6 \times 10^6$ & $15 \times 10^6$ \\
        density             & $0.018$ & $0.018$ & $0.024$ & $0.025$ & $0.026$ & $0.029$ \\
        max                 & $1120$ & $5.6 \times 10^{4}$ & $6.0 \times 10^8$ & $1.9 \times 10^{12}$ & $5.7 \times 10^{17}$ & $2.0 \times 10^{25}$ \\
        \bottomrule
    \end{tabular}
    \caption{
        Features of the $\funspace_\weight$ basis we have constructed, in its  coproduct tensor representation. The first row represents the weight
        $\weight$, the second row gives the dimension of the space \brk{including MZVs}, the third, fourth and fifth rows show the amount of non-zero
        entries, densities and maximum entry values of the coproduct tensors, respectively.  }
    \label{tab:coprodsSize}
\end{table}

Now let us turn to the construction of a particular basis $\brc{\stf\supbrk{\weight}_j}_{j \in J_{\weight}}$ of the $\funspace_{\weight}$ space
of the DS functions of weight $\weight \ge 1$, where the elements of the basis are labeled by a list $J_{\weight} \defas \brc{1, 2, \ldots, \abs{J_{\weight}}}$.
In this section we construct the $\funspace_{\weight}$ in the so-called coproduct form, then further refine it in \secref{ssec:branchCutCond}, and
finally promote it to the full functional form in \secref{ssec:funs}. But before we do this, we first need to properly
introduce the initial conditions and our notation for coproducts.

The starting point of our recursive construction is the lowest weight space $\funspace_1$, which is two dimensional:
$\dim_{\Rationals}\brk{\funspace_1} \equiv \abs{J_1} = 2$, and is spanned by the following two logarithms:
\begin{align}
    \stf\supbrk{1}_1 \defas G\brk{\tfrac1y; x} = \log\brk{1 - x y}= - \sum_{n \ge 1} \tfrac1n \brk{x y}^n,
    \label{eq:stf11}
    \\
    \stf\supbrk{1}_2 \defas G\brk{1; y} + G\brk{0; x} = \log\bigbrk{x \brk{1 - y}}.
    \label{eq:stf12}
\end{align}
Note, that around $x = 0$ the first function $\stf\supbrk{1}_1$ has a well-defined Taylor expansion, while the second function
$\stf\supbrk{1}_2$ is logarithmically divergent. In \secref{sec:hex} such power-and-log expansions of the elements of $\funspace$
will become our main tool for calculating amplitudes in the DS limit.

Next, as was outlined in \secref{ssec:notation}, the $\brc{\weight - 1, 1}$ coproduct component of a weight $\weight$ DS function $\stf\supbrk{\weight}_i \in \funspace_{\weight}$ can be expressed
as a $\Rationals$-linear combination of certain $\otimes$-products with weight $\brk{\weight - 1}$ DS functions $\stf\supbrk{\weight - 1}_j \in \funspace_{\weight - 1}$
in their left entries and logarithms of the $\alphabet$ alphabet elements in the right. As was first understood in \cite{Dixon:2013eka} polylogarithmic functions are more economically expressed in terms of this coproduct component, and as advocated in \cite{Dixon:2016nkn}, see also \cite{Caron-Huot:2019bsq}, it is more efficient to encode this component using a single object
$\cc{\weight}{i j \alpha} \in \Rationals^{\abs{J_{\weight}} \times \abs{J_{\weight - 1}} \times \abs{\alphabet}}$,
which we refer to as the coproduct tensor, namely
\begin{align}
    \cpr_{\weight - 1, 1}\bigbrk{\stf_i\supbrk{\weight}}
    = \sum_{j, \alpha}
    \cc{\weight}{i j \alpha} \,
    \stf\supbrk{\weight - 1}_j \otimes \log\brk{\alpha},
    \label{eq:cpr1}
\end{align}
where $i \in J_{\weight}$ and we have omitted the summation ranges $j \in J_{\weight - 1}$ and $\alpha \in \alphabet$ for simplicity.
It follows from the coassociativity of the coproduct $\Delta$, that its $\brc{\weight - 2, 1, 1}$ component is then given by an
inner product of two coproduct tensors at weight $\weight$ and $\weight - 1$:
\begin{align}
    \cpr_{\weight - 2, 1, 1}\bigbrk{\stf_i\supbrk{\weight}}
    = \sum_{\substack{j, k \\ \alpha, \beta}}
    \cc{\weight}{i j \alpha} \,
    \cc{\weight - 1}{j k \beta} \,
    \stf\supbrk{\weight - 2}_k \otimes \log\brk{\beta} \otimes \log\brk{\alpha}.
    \label{eq:cpr11}
\end{align}

Now we are ready to discuss the recursive step, i.e. the construction of the $\funspace_{\weight}$ out of the already known $\funspace_{\weight - 1}$
space at previous weight. Following the coproduct bootstrap method, we formulate a set of linear constraints on the coproducts, whose nullspace
defines the $\funspace_{\weight}$ space.
To do that, we start with a tensor product $\cpr_{\weight - 2, 1} \brk{\funspace_{\weight - 1}} \otimes \alphabet$ of the $\brc{\weight - 2, 1}$
coproduct component of the whole $\funspace_{\weight - 1}$ space, encoded in the $\cc{\weight - 1}{j k \beta}$ tensor, and another copy of the
$\alphabet$ alphabet. Using the natural inclusion
$\cpr_{\weight - 2, 1} \brk{\funspace_{\weight - 1}} \otimes \alphabet \subset \funspace_{\weight - 2} \otimes \alphabet \otimes \alphabet$,
we impose a set $\intSet \defas \brc{1, \ldots, \abs{\intSet}}$ of homogeneous linear conditions on the last two entries of the coproduct
via a map $\intTensor{} : \alphabet \otimes \alphabet \to \Rationals^{\abs{\intSet}}$ that acts on the tensor space as follows:
\begin{align}
    \funspace_{\weight - 1} \otimes \alphabet
    \xrightarrow[]{\cpr_{\weight - 2, 1} \otimes \> \id}
    \cpr_{\weight - 2, 1} \brk{\funspace_{\weight - 1}} \otimes \alphabet
    \xrightarrow[]{\id \otimes \> \intTensor{}}
    \funspace_{\weight - 2} \otimes \Rationals^{\abs{\intSet}},
    \label{eq:intCondMap}
\end{align}
and look for its kernel. In our implementation $\intTensor{}$ contains the integrability as well as the extended Steinmann conditions shown
in eqs. \eqref{eq:integrability} and \eqref{eq:steinmann}, so that $\abs{\intSet} = 8$ in our case.
In practice, the map $\intTensor{}$ is represented as a tensor
$\intTensor{\alpha \beta s} \in \Rationals^{\abs{\alphabet} \times \abs{\alphabet} \times \abs{\intSet}}$, whose explicit values are stated
in eq. \eqref{eq:intTensorDef}.
Contraction of the $\intTensor{\alpha \beta s}$ tensor together with the coproduct tensor $\cc{\weight - 1}{j k \beta}$ gives a realization of the
$M \defas \sbrk{\id \otimes \> \intTensor{}} \circ \sbrk{\cpr_{\weight - 2, 1} \otimes \id} $
map from eq. \eqref{eq:intCondMap}, whose tensorial representation \brk{after some transpositions} can easily be spelled out:
\begin{align}
    M_{k s j \alpha} \defas \sum_{\beta} \cc{\weight - 1}{j k \beta} \, \intTensor{\alpha \beta s}.
    \label{eq:Mdef}
\end{align}
A more explicit derivation of this definition is given in \appref{app:tensorCoprod}. This map $M$ encodes the basic linear constraints needed for
determining the $\funspace_{\weight}$ space.

Now, let $\cct{\weight}{\tilde{i} \brk{j \alpha}} \in \Rationals^{\abs{\tilde{J}_{\weight}} \times \brk{\abs{J_{\weight - 1}} \cdot \abs{\alphabet}}}$
denote a basis of the $\kernel\brk{M_{\brk{k s} \brk{j \alpha}}}$ nullspace:
\begin{align}
    \sum_{\brk{j \alpha}} \,
    M_{\brk{k s} \brk{j \alpha}} \,
    \cct{\weight}{\tilde{i} \brk{j \alpha}}
    = 0, \quad
    \text{for each }
    k \in J_{\weight - 1}, \>
    s \in \intSet, \>
    \tilde{i} \in \tilde{J}_{\weight},
    \label{eq:nullspaceBasis}
\end{align}
where the list $\tilde{J}_{\weight} \defas \brc{1, \ldots, \abs{\tilde{J}_{\weight}}}$, for the moment, labels the basis elements, while the brackets
$\brk{j \alpha}$ and $\brk{k s}$ denote vectorization of the corresponding tensor indices, which label bases of the leftmost and the rightmost spaces
in eq.  \eqref{eq:intCondMap} respectively.
To further reduce the nullspace $\kernel\brk{M_{\brk{\alpha j} \brk{s k}}}$ and obtain the entire $\funspace_{\weight}$ space we exploit
the additional analyticity constraints shown in eq. \eqref{eq:brcond}.
Each of these 3 branch cut conditions always evaluate to just MZVs, hence they produce $3 \times \abs{\Zetas_{\weight - 1}}$ additional linear
equations at weight $\weight$. Resolving these conditions further reduces the list of basis elements $\tilde{J}_\weight \to J_\weight$, and leaves
us with the final form of the $\cc{\weight}{i j \alpha}$ coproduct tensors. Their properties with respect to the weight $\weight$ are summarized in \tabref{tab:coprodsSize}. Note,
in particular, that the produced coproduct tensors tend to be very sparse and their densities never exceed a few percent. We present the coproduct
tensors for weight $\weight \le 13$ in the ancillary file \ancillary{coproducts-w2-13.m}.

In practice,
the natural $\GL\brk{\abs{\tilde{J}_{\weight}}}$ freedom of choosing a particular version of the coproduct tensors \brk{prior to imposing the branch cut
conditions} $\cct{\weight}{\tilde{i} \brk{j \alpha}}$ can be used to improve the efficiency of the computer implementation.
There are, of course, many possible metrics for optimization: we can look for the most sparse tensors in the output, the fastest overall execution
time, the lowest values of the $\lVert \cdot \rVert_\infty$ norm, or some heuristic mix of those.  The main computational challenge is the Gaussian
elimination in the field of rationals $\Rationals$ required for the nullspace determination.  As the dimensions of the equation matrix $M_{\brk{k s}
\brk{j \alpha}}$ grow exponentially with increasing weight $\weight$, so do the denominators and numerators in its entries. A very special care is
therefore needed in order to perform the Gaussian elimination at higher weights.
Building on the ideas of \cite{Drummond:2014ffa}, we tested several possible optimization strategies, the most promising of which are presented in
\tabref{tab:numerics}.
Our best method of computation consists of reversing the order of vectors of the \code{NullSpace}'s output and application of
the
Lenstra–Lenstra–Lovasz (LLL) reduction via \mathematica's \code{LatticeReduce} command. This way we were able to significantly reduce the sizes of the integer entries in the
produced coproduct tensors $\cc{\weight}{j k \beta}$, as reflected in \tabref{tab:numerics}. At higher weights $\weight \ge 11$ we also made
use of the \code{Spasm} library~\cite{spasm} for performing the row reduction over finite fields and subsequent rational reconstruction.

\begin{table}
    \centering
    \begin{tabular}{c|cccc}
        \toprule
        weight  & A     & B     & C     & D \\
        \midrule
        2       & 1     & 1     & 1     & 1 \\
        3       & 2     & 2     & 1     & 1 \\
        4       & 2     & 2     & 2     & 2 \\
        5       & 20    & 20    & 10    & 7 \\
        6       & 40    & 20    & 60    & 105 \\
        7       & 560   & 420   & 2100  & 18366 \\
        8       & 1120  & 14850 & 5950  & 227673167 \\
        \bottomrule
    \end{tabular}
    \caption{
        The sizes of maximal entries of the coproduct tensors $\cc{\weight}{i j \alpha}$ at weight $\weight$ in the four computational setups
        that we used. The version A, our best choice so far, reorders the rows of the tensors and uses the LLL reduction; the version B in addition to that also rescales the
        rows of the tensors in order to make them integer-valued and only then applies the LLL; the version C only uses the LLL without any reordering
        of the rows; version D does not modify the tensors in any way and serves as a baseline for numeric optimizations.
    }
    \label{tab:numerics}
\end{table}

\subsection{Promoting coproducts to functions}
\label{ssec:funs}
Now everything is ready for construction of the full function space $\funspace$ that is used to determine the MHV and NMHV six-gluon amplitudes in the
DS limit. We provide two equivalent representations of the elements of $\funspace$: one in terms of MPLs and the other one in terms of power-and-log
series expansions. The latter is less computationally demanding and instrumental for the Wilson loop OPE resummation as explained in \secref{ssec:wlope}.

\begin{figure}[t]
    \centering
    \includegraphics[align=c]{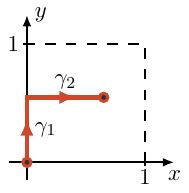}
    \caption{
        Integration path $\gamma \defas \gamma_1 \gamma_2$ connecting the base point of integration $\brc{0, 0}$ and a general kinematical point
        $\brc{x, y}$ inside the unit square.
    }
    \label{fig:intPath}
\end{figure}

Our procedure essentially boils down to iterative integration of the differentials~\ref{dFPhi} defined by the coproducts from eq. \eqref{eq:cpr1}
along a fixed path $\intpath$ shown in \figref{fig:intPath}. The path $\intpath \defas \intpath_1 \intpath_2$ connects the base point of integration
$\brc{x, y} = \brc{0, 0}$, at which we set the integration constants to $0$, with a general kinematical point $\brc{x, y}$ inside the unit square
$0 < x, y < 1$.
Our choice of the integration path $\gamma$ dictates a special representation of the DS functions $\stf\supbrk{\weight}_j \in \funspace_{\weight}$ in
terms of MPLs, see for example~\cite{Brown:2009qja} or~\cite{Panzer:2015ida},
\begin{align}
    \stf\supbrk{\weight}_{j} = \sum\limits_{\vec{X}, \vec{Y}} g_{j \vec{X} \vec{Y}} \> G\brk{\vec{X}; x} \, G\brk{\vec{Y}; y},
    \quad
    g_{j \vec{X} \vec{Y}} \in \Zetas^{\abs{J_{\weight}} \times 3^\weight \times 2^\weight},
    \quad
    \bigabs{g_{j \vec{X} \vec{Y}}} + \bigabs{\vec{X}} + \bigabs{\vec{Y}} = \weight,
    \label{eq:stfG}
\end{align}
where the lists \brk{which are also referred to as `words'} of arguments $\vec{Y}$ and $\vec{X}$ are drawn from the $\vec{Y} \in \brc{0, 1}^{\bigcdot}$
and $\vec{X} \in \brc{0, 1, \tfrac1y}^{\bigcdot}$ sets.
Here by $\Sigma^{\bigcdot}$ we mean the set of all words made out of some finite set $\Sigma$, in other words it is a disjoint union
$\Sigma^{\bigcdot} \defas \emptyword \cupdot \Sigma \cupdot \Sigma^2 \cupdot \ldots$,
where $\emptyword$ is an empty word, while any other word $\vec{X}$ of length $\bigabs{\vec{X}}$ lies in the $\abs{\vec{X}}^{\text{th}}$ component: $\vec{X} \in \Sigma^{\abs{\vec{X}}}$.
The lengths of these two lists and the transcendental weight of the MZV-valued coefficients, denoted as $\bigabs{\vec{X}}$, $\bigabs{\vec{Y}}$, and
$\bigabs{g_{j \vec{X} \vec{Y}}}$ respectively, add up to the transcendental weight $\weight$ of the function for each term in the sum.

\begin{table}
    \centering
    \begin{tabular}{c|ccccc}
        \toprule
        alphabet $\alphabet$& $x$               & $y$               & $1 - x$               & $1 - y$               & $1 - x y$ \\
        \midrule
        $\intpath_1$        & $0$               & $\dlog\brk{y}$    & $0$                   & $\dlog\brk{y - 1}$    & $0$ \\
        $\intpath_2$        & $\dlog\brk{x}$    & $0$               & $\dlog\brk{x - 1}$    & $0$                   & $\dlog\brk{x - \tfrac1y}$ \\
        \bottomrule
    \end{tabular}
    \caption{
        Pull-backs of the $\dlog$-forms along the two components of the integration path shown in
        \protect\figref{fig:intPath}. These $\dlog$-forms act as the integration kernels in eqs. \protect\eqref{eq:intO1, eq:intO2, eq:intO3}.
        Note, that along $\intpath_1$ the integration variable is $y$, and along $\intpath_2$ the integration variable is $x$, while $y$ remains
        constant.
    }
    \label{tab:dlogs}
\end{table}

The integral along $\intpath$ of the $\brc{\weight - 1, 1}$ coproduct component \eqref{eq:cpr1} or, equivalently, of the differential of a given
DS function splits into two terms corresponding to the two parts of the integration path.
The pull-backs of the $\dlog$-forms shown in \tabref{tab:dlogs} make sure that both of these integrations preserve the general form of the
\eqref{eq:stfG} representation: the integral along the $\intpath_1$ only modifies the $g_{j \emptyword \vec{Y}}
G\brk{\vec{Y}; y}$ terms in that sum, while the $\intpath_2$ integration changes only the $G\brk{\vec{X}; x}$ factor.
To quantitatively describe the second integral along the $\intpath_2$ path we introduce the following three linear operators:
\begin{alignat}{2}
    \op_1\brk{G\brk{\vec{X}; x}} &\defas \int_0^x G\brk{\vec{X}; t} \> \dlog{\brk{t}} &= G\brk{0, \vec{X}; x},
    \label{eq:intO1}
    \\
    \op_2\brk{G\brk{\vec{X}; x}} &\defas \int_0^x G\brk{\vec{X}; t} \> \dlog{\brk{t - 1}} &= G\brk{1, \vec{X}; x},
    \label{eq:intO2}
    \\
    \op_3\brk{G\brk{\vec{X}; x}} &\defas \int_0^x G\brk{\vec{X}; t} \> \dlog{\brk{t - \tfrac1y}} &= G\brk{\tfrac1y, \vec{X}; x}.
    \label{eq:intO3}
\end{alignat}
It is evident that these operators just prepend the corresponding letter to the list of arguments. Later in eqs. \eqref{eq:intO1exp, eq:intO2exp,
eq:intO3exp} we will show how the same integral operators act on the series expansions of MPLs, which turn out to be more useful objects in practice.

In this manner, we have succeeded in promoting the coproduct representation of our DS functions to explicit expressions in terms of MPLs through weight 12. Due to size restrictions, in the ancillary file \ancillary{ds-w1-10.m} we provide these MPL representations through weight 10 and list the weight $\weight \le 3$ functions explicitly below.\footnote{Readers interested in the $p=11,12$ expressions are welcome to contact the authors.}
In terms of the following compact notation,
\begin{align}
    l_x \defas \log\brk{x}, \qquad \Gx_{\vec{X}} \defas G\brk{\vec{X}; x}.
    \label{eq:notation}
\end{align}
which we will adopt from this point on, the weight $\weight = 1$ functionsof eqs. \eqref{eq:stf11} and \eqref{eq:stf12} read
\begin{align}
    \stf\supbrk{1}_1 &= l_{1-x y}, \>
    \stf\supbrk{1}_2 = l_{x (1-y)}.
    \label{eq:stf1}
\end{align}
Then, there are five functions at weight $\weight = 2$,
\begin{align}
    \stf\supbrk{2}_1 &= \Gx_{0 \yinv}, \>
    \stf\supbrk{2}_2 = \Gx_{0 \yinv}-\Gx_{1 \yinv}-\Gx_{0 1}+l_{1-x} l_{x (1-y)}, \>
    \nn\\
    \stf\supbrk{2}_3 &= \tfrac{1}{2} l_{x (1-y)}^2, \>
    \stf\supbrk{2}_4 = l_{x (1-y)} l_{1-x y}-l_{1-x y}^2, \>
    \stf\supbrk{2}_5 = \zeta_2\,, \>
    \label{eq:stf2}
\end{align}
and 12 functions at $\weight = 3$,
\begin{align}
    \stf\supbrk{3}_1 &= \Gx_{\yinv 0 \yinv}, \>
    \stf\supbrk{3}_2 = \zeta_2 l_{1-x y}, \>
    \stf\supbrk{3}_3 = \zeta_2 l_{x (1-y)}, \>
    \stf\supbrk{3}_4 = \Gx_{0 0 \yinv}, \>
    \nn\\
    \stf\supbrk{3}_5 &= \Gx_{0 1} l_{x (1-y)}+\Gx_{0 0 \yinv}-\Gx_{0 1 \yinv}-2 \Gx_{0 0 1}, \>
    \stf\supbrk{3}_6 = \tfrac{1}{6} l_{x (1-y)}^3, \>
    \nn\\
    \stf\supbrk{3}_7 &=
    \tfrac{1}{2} l_{1-x y} l_{x (1-y)}^2
    + l_{x (1-y)} \sbrk{
        -\Gx_{0 \yinv}-\Gx_{\yinv 1}
    } +\Gx_{0 0 \yinv}+\Gx_{0 \yinv 1}+\Gx_{\yinv 0 1}-\Gx_{\yinv 0 \yinv}+\Gx_{\yinv 1 \yinv}, \>
    \nn\\
    \stf\supbrk{3}_8 &= \Gx_{0 \yinv} l_{x (1-y)}-2 \Gx_{0 \yinv \yinv}, \>
    \nn\\
    \stf\supbrk{3}_9 &=
    \tfrac{1}{2} l_{1-x} l_{x (1-y)}^2
    + l_{x (1-y)} \sbrk{
        -\Gx_{0 1}-\tfrac{1}{2} l_{1-x}^2
    }
    -\Gx_{1 0 \yinv}+\Gx_{1 1 \yinv}+\Gx_{0 0 1}+\Gx_{0 1 1}+\Gx_{1 0 1}+\zeta_2 l_{1-x}, \>
    \nn\\
    \stf\supbrk{3}_{10} &= l_{x (1-y)} \sbrk{
        \Gx_{0 \yinv}-\Gx_{1 \yinv}
    }
    -\Gx_{0 0 \yinv}+\Gx_{0 1 \yinv}-2 \Gx_{0 \yinv \yinv}+2 \Gx_{1 \yinv \yinv}, \>
    \nn\\
    \stf\supbrk{3}_{11} &=
    l_{x (1-y)} \sbrk{
        -\Gx_{1 \yinv}-\Gx_{\yinv 1}+l_{1-x y}^2
    }
    +\Gx_{0 1 \yinv}+\Gx_{0 \yinv 1}-2 \Gx_{0 \yinv \yinv}+2 \Gx_{1 \yinv \yinv}+\Gx_{\yinv 0 1}-\Gx_{\yinv 0 \yinv}+\Gx_{\yinv 1 \yinv}
    -\tfrac{2}{3} l_{1-x y}^3, \>
    \nn\\
    \stf\supbrk{3}_{12} &= \zeta_3.
    \label{eq:stf3}
\end{align}

As was anticipated at the end of \secref{ssec:DS_general}, we see the low-weight MZV constants $\zeta_2$ and $\zeta_3$ are independent elements of
the $\funspace$ space, the same holds for other MZVs from eq. \eqref{eq:mzvbasis} at higher weights.

Before concluding this subsection, let us also make a technical remark regarding the efficient representation of MPLs for computer algebra
manipulations \cite{Blumlein:2009cf}: It proves useful to store the $\vec{Y}$ lists of arguments of MPLs from eq. \eqref{eq:stfG} as base 2 \brk{or binary} numbers, and
encode the $\vec{X}$ lists
as base 3 \brk{or ternary} numbers via a $\tfrac1y \to 2$ replacement, e.g.
$G\brk{1, 0, \tfrac1y; x} \mapsto G\brk{102_3; x}$.
It allows us to represent the action of the $\brc{\op_1, \op_2, \op_3}$ integration operators arithmetically as addition of certain \brk{weight-dependent}
numbers, e.g.
$\op_3\brk{G\brk{1, 0, \tfrac1y; x}} = G\brk{\tfrac1y, 1, 0, \tfrac1y; x} \mapsto G\brk{102_3 + 2000_3; x}$.
Also note that as a consequence of the branch cut condition of eq. \eqref{eq:brcond}, $f^y = 0$ as $y \to 0$, the first integral along the $\intpath_1$
effectively only increases the powers of $l_{1 - y} \defas \log\brk{1 - y}$, since the $\vec{Y}$ lists in eq. \eqref{eq:stfG} are forced to be free of any zeros inside
of them. Therefore we only need to keep track of its overall length $\abs{\vec{Y}}$.
We have found that these tweaks greatly to reduce the memory and storage usage of our computer implementation.

\subsection{Promoting coproducts to expansions}
\label{ssec:expansions}
For every function inside our $\funspace$ space, apart from its MPL representation $\stf$ we also construct its $x \to 0$ expansion, which we shall denote as $\estf$ in order to avoid confusion \brk{note that we have also dropped the weight index to avoid clutter}.
In the next section we will use these expansions $\estf$ to resum the Wilson loop OPE state sum and obtain predictions for
the NMHV six gluon amplitude at high loop orders, so before that we need to discuss the relevant features of our construction.

In fact, direct integration of the $x \to 0$ expansions $\estf$ turns out to be easier than expanding the MPL representations we obtained in the previous subsection,
especially at higher weights $\weight \ge 10$. This way at each step of the algorithm we only have to deal with $x \to 0$ power-and-log expansions
of MPLs, which are directly applied to fix the coefficient of the ansatz in the NMHV amplitude bootstrap problem.
We will consider the $x \to 0$ expansions $\estf$ of the following mixed form:
\begin{align}
    \estf
    =
    \sum_{k = 0}^N \sum_{l = 0}^{k - 1} \sum_{m, n} \>
    q_{k l m n}
    x^k y^l \bigbrk{\log\brk{x}}^m \bigbrk{\log\brk{\ym}}^n
    + \sum\limits_{\tilde{X}} g_{\tilde{X}} \> G\brk{\tilde{X}; x}
    \label{eq:stfExp}
\end{align}
where the list of arguments of MPLs $\tilde{X}$ is drawn from a reduced set $\tilde{X} \in \brc{0, \tfrac1y}^{\bigcdot}$,
and the coefficients
$q_{k l m n} \in \Zetas^{N \times \brk{N - 1} \times \weight \times \weight}$ and
$g_{\tilde{X}} \in \Zetas^{2^\weight}$ are MZV-valued, so that the overall weight of the RHS in eq. \eqref{eq:stfExp}
does not exceed $\weight$, while $\bigabs{g_{\tilde{X}}} + \bigabs{\tilde{X}} = \weight$.
The summation range $0 \le l < k \le N$ in the first sum of eq. \eqref{eq:stfExp} indicates that the MPLs that contribute to it
have at least one ``$1$'' among their arguments. In contrast, MPLs from the second sum of eq. \eqref{eq:stfExp}, when expanded,
only contain equal powers of both $x$ and $y$ \brk{apart from simple logarithms}. Therefore, we refer to the first sum in eq. \eqref{eq:stfExp} as the
``non-diagonal part'' of the expansion and denote the corresponding linear space of monomials by $\ndiagSt$, while the second sum will be called the
``diagonal part'', whose linear space of MPLs \brk{or $x \to 0$ expansions} is denoted by $\diagSt$.
Hence an expansion $\estf$ belongs to a direct sum of these two parts: $\estf = \ndiagSt \oplus \diagSt$ and so does the NMHV
amplitude that we are after.

Alongside the $\brc{\weight - 1, 1}$ coproduct components discussed in \secref{ssec:coprods}, our main computational tool are the integration operators
introduced in eqs. \eqref{eq:intO1, eq:intO2, eq:intO3}. Their action on the non-diagonal $\ndiagSt$ part of the expansions $\estf$ is given by the
following formulas:
\begin{align}
    \left.\op_1\brk{\estf}\right|_\cN &\defas \sum_{k, l, m, n} q_{k l m n} \> x^k y^l \blogym^n \, \sum_{i = 0}^m
    \brk{-1}^{m - i} \> \frac{m!}{i!} \, k^{-\brk{m - i + 1}} \> \blogx^i
    \label{eq:intO1exp}
    \\
    \left.\op'_2\brk{\estf}\right|_\cN &\defas -\sum_{k, l, m, n} q_{k l m n} \> y^l \blogx^m \blogym^n \, \sum_{i = k + 1}^N x^i
    \label{eq:intO2exp}
    \\
    \left.\op'_3\brk{\estf}\right|_\cN &\defas -\sum_{k, l, m, n} q_{k l m n} \> y^{l - k} \blogx^m \blogym^n \, \sum_{i = k + 1}^N \brk{x y}^i
    \label{eq:intO3exp}
\end{align}
where the partial operators $\brc{\op'_{2}, \op'_3}$ are related to the full integration operators $\brc{\op_{2}, \op_3}$ as $\op_2 = \op_1 \circ \op'_2$ and
$\op_3 = \op_1 \circ \op'_3$. Note, that the rule \eqref{eq:intO1exp} needs an additional definition in case of a pure $\logx$ monomial:
\begin{align}
    \op_1\brk{\blogx^m} \defas \frac{1}{m + 1} \blogx^{m + 1},
\end{align}
in accordance with eq. \eqref{eq:GdefLog}.

\begin{table}
    \centering
    \begin{tabular}{c|cc}
        \toprule
                    & $\ndiagSt$          & $\diagSt$\\
        \midrule
        $\op_1$     & $\ndiagSt$    & $\diagSt$\\
        $\op'_2$    & $\ndiagSt$    & $\ndiagSt$\\
        $\op'_3$    & $\ndiagSt$    & $\diagSt$\\
        \bottomrule
    \end{tabular}
    \caption{
        Action of the integration operators $\brc{\op_1, \op'_2, \op'_3}$ on the non-diagonal $\ndiagSt$ and diagonal $\diagSt$ parts of the
        expansions.
        None of the operators maps the non-diagonal part $\ndiagSt$ into the diagonal $\diagSt$, and the $\op_1$ and $\op'_3$ operators even preserve
        $\diagSt$. These properties allow us to keep the diagonal $\diagSt$ part of the expansion in eq. \protect\eqref{eq:stfExp} in the full MPL
        form and only expand the non-diagonal $\ndiagSt$ part.
    }
    \label{tab:expOper}
\end{table}

The action of $\brc{\op_1, \op'_2, \op'_3}$ is summarized in \tabref{tab:expOper} in terms of the non-diagonal and diagonal parts $\ndiagSt$ and $\diagSt$.
A crucial observation about these operators is that they do not increase the power of $y^l$ independently from $x^k$ or, in other words,
none of them act as $\ndiagSt \to \diagSt$, which is, of course, just a consequence of the eqs. \eqref{eq:intO1, eq:intO2, eq:intO3}.
This feature allows us to work with the diagonal part $\diagSt$ of any expansion $\estf$ separately keeping it in the full unexpanded MPL form.

\begin{table}
    \centering
    \begin{tabular}{c|ccc ccc c}
        \toprule
        weight $\weight$        & $\le 6$ & $7$ & $8$ & $9$ & $10$ & $11$ & $12$\\
        order $\Nopt$  & $\le 4$ & $5$ & $7$ & $9$ & $13$ & $18$ & $25$\\
        \bottomrule
    \end{tabular}
    \caption{
        Optimal expansion order $\Nopt$, corresponding to the minimal value of $N$ in eq. \protect\eqref{eq:stfExp}, such that the series expansions $\estf$ of all DS functions at a given weight $\weight$ become linearly independent.
    }
    \label{tab:order}
\end{table}

Finally, let us note an important technical detail of this approach: the maximal expansion order $N$ needed for the high-weight functions
has to be set in advance for the weight $\weight = 1$ functions, and cannot be altered in between.
We define the optimal expansion order $N\brk{\weight} = \Nopt\brk{\weight}$ to be
the minimal value of $N$, for which the expansions of the weight $\weight$ functions we obtain in this way, are linearly independent. Our empirical findings
of the value of $\Nopt\brk{\weight}$ are summarized in \tabref{tab:order}. We emphasize that had we chosen to also expand the diagonal part $\diagSt$,
the optimal expansion order would drastically increase. The computational gain obtained thus justifies our hybrid strategy of only expanding the
non-diagonal $\ndiagSt$ part in eq.~\eqref{eq:stfExp}.

\section{The Wilson loop OPE and the collinear limit}
\label{sec:hex}

In the previous sections, we essentially described how to construct an ansatz for the six-gluon amplitude in the DS limit. Here, we will show how to find a unique solution for this ansatz, by exploiting independent information for the amplitude in the collinear boundary of the DS limit, furnished by the Wilson Loop OPE  approach~\cite{Basso:2013vsa, Basso:2013aha, Basso:2014koa, Basso:2014nra, Belitsky:2014sla, Belitsky:2014lta, Basso:2014hfa,
Basso:2015rta, Basso:2015uxa, Belitsky:2016vyq}.

While our space of functions contains both the MHV and NMHV amplitudes, in what follows we will be focusing on the latter, which in the literature is known to lower loop order~\cite{Caron-Huot:2019vjl}. After reviewing another amplitude normalization and set of kinematic variables that will be convenient for our purposes, in subsection~\ref{ssec:wlopeIntro} we analyze the behavior of the rational, $R$-invariant part of the NMHV amplitude in the two parity-conjugate DS limits. Choosing one of the two limits, and the so-called (1111) rational component of the amplitude, subsection~\ref{ssec:wlope} then discusses in detail how to obtain predictions from the Wilson loop OPE, and match them against our ansatz. The reader interested in the final result, may jump to \secref{sec:results}.

\subsection{Wilson loop normalization, variables and $R$-invariants in the DS limit}
\label{ssec:wlopeIntro}

In order to describe the Wilson loop dual to the amplitude, it will be convenient to change the BDS-like normalization \eqref{eq:cEEDef}, redistributing a known factor between the finite and infrared-divergent part of the latter. That is, in this section we will be considering the NMHV framed Wilson loop, $\Wmhv$, which is related to $\Enmhv$ by
\begin{align}
    \frac{\Enmhv}{\Wnmhv}&=\> \exp\bigbrk{-\tfrac14 \gcusp \expX}.
    \label{eq:EmhvDef}
\end{align}
In the rightmost factor in this equation $\tfrac14 \gcusp = g^2 - 2 \zeta_2 g^4 + \order{g^6}$ is the cusp anomalous dimension, a quantity known to all loops~\cite{Beisert:2006ez}, and
\begin{align}
    \expX &\defas X - \Emhv\supbrk{1}
    \label{eq:expXdef}
    \nn\\
    X &= -\Li_2\brk{1 - u} - \Li_2\brk{1 - v} + \Li_2\brk{w}
    \nn\\
    &\quad+ \log\brk{1 - w}^2 - \log\brk{1 - w}\log\brk{\tfrac{v}{u}} - \log\brk{u}\log\brk{v}
    + \tfrac{\pi^2}{6},
    \\
    \Emhv\supbrk{1} &= \Li_2\brk{1 - \tfrac1u} + \Li_2\brk{1 - \tfrac1v} + \Li_2\brk{1 - \tfrac1w}.
\end{align}
As we mentioned in~\secref{sec:review}, at each loop order $L$, $\Emhv\supbrk{L}$ is believed to be a pure transcendental function, and particularly in the DS limit it
should lie in the $\funspace$ space. On the other hand, as shown in eq.\eqref{eq:EnmhvDef} $\Enmhv$ is a combination of pure functions and rational $R$-invariants. The latter are polynomials in the Grassmann variables encoding the supersymmetry of the theory, and in this work we will particularly focus on their $\brk{1111}$ component, first studied from the point of view of the Wilson loop OPE in~\cite{Basso:2013aha}.
%

%
Following the latter reference, we will also switch to another choice of kinematic variables that is convenient for describing both the collinear and DS limits of the hexagonal Wilson loop. It consists of a triple
$\brc{S, F, T}$, whose relation to the $\brc{u, v, w}$ cross-ratios reads:
\begin{align}
    u = \frac{F}{F + F S^2 + S T + F^2 S T + F T^2},
    \quad
    v = \frac{S^2}{T^2} \, u \, w,
    \quad
    w = \frac{T^2}{1 + T^2}.
    \label{eq:crWlope}
\end{align}
In these variables, for the NMHV $\brk{1111}$ component of interest the $R$-invariants reduce to \cite{Caron-Huot:2019vjl}
\begin{align}
    \brk{1} &\to 0,
    \nn\\
    \brk{2} &\to \frac{F^3 T}{
        \brk{S + F T}
        \brk{F + S T + F T^2}
        \brk{F + F S^2 + S T + F^2 S T + F T^2}
    },
    \nn\\
    \brk{3} &\to \frac{1}{
        \brk{1 + T^2}
        \brk{1 + F S T + T^2}
    },
    \label{eq:RinvsDS}
    \\
    \brk{4} &\to \frac{S}{S + F T},
    \nn\\
    \brk{5} &\to \frac{T (F S + T)^3}{
        F
        \brk{1 + F S T + T^2}
        \brk{F + F S^2 + S T + F^2 S T + F T^2}
    },
    \nn\\
    \brk{6} &\to \frac{T^4}{
        F
        \brk{1 + T^2}
        \brk{F + S T + F T^2}
    }.
    \nn
\end{align}

Now let's see what are the simplifications to the above formulas that follow from the double scaling kinematics.
As was shown in \secref{ssec:DS_general}, there are really two DS limits, which in  the OPE variables~\eqref{eq:crWlope} correspond to:
\begin{alignat}{3}
    &\text{double scaling limit 1 \brk{DS1}:} \quad &&\brc{T, F^{-1}} \to 0,  \quad &&T F \; \text{fixed,}
    \label{eq:DS1def}
    \\
    &\text{double scaling limit 2 \brk{DS2}:} \quad &&\brc{T, F} \to 0,       \quad &&T F^{-1} \; \text{fixed.}
    \label{eq:DS2def}
\end{alignat}
They are related by the parity transformation, in these variables translates to a simple $F \to F^{-1}$ replacement.
In these limits, we may also relate the $\brc{x, y}$ variables of eq. \eqref{eq:xytou} with respect to the surviving OPE variables as
$\brc{x, y} \big|_{\text{DS1}} = \brc{- \frac{T F}{S}, 1 + \frac{1}{T F S}}$
and
\begin{align}
    \brc{x, y}\big|_{\text{DS2}} =
    \Bigbrc{- \frac{T}{S F}, 1 + \frac{F}{T S}}.
    \label{eq:xyDS2}
\end{align}
From now on will will thus switch back to our familiar $\brc{x, y}$ variables.

Also note that the cross-ratios eq. \eqref{eq:crWlope} are parity-even functions, so they reduce in both versions of the DS limit to
the same values,
$\brc{u, v, w} \to \Bigbrc{\frac{x \brk{1 - y}}{1 - x y}, \frac{1}{1 - x y}, 0}$.
Therefore, the function $\expX$ responsible for the BDS-like normalization and defined in eq. \eqref{eq:expXdef} in both of the DS
limits reduces to
\begin{align}
    \expX &\xrightarrow[\text{DS}]{} \tfrac{\pi^2}{3} + \tfrac12 \log\brk{x \brk{1 - y}} + \tfrac12 \log\brk{w}^2.
\end{align}

On the other hand, the $\brk{1111}$ components of the $R$-invariants in the two versions of the double scaling limit differ significantly.
Using the definitions from eqs. \eqref{eq:RinvsDS} and the limits from eqs. \eqref{eq:DS1def} and \eqref{eq:DS2def} we see that
\begin{align}
    \brc{\brk{1}, \ldots, \brk{6}}
    \to
    \begin{cases}
        \brc{0, -\frac{x^2 \brk{1 - y}}{\brk{1 - x} \brk{1 - x y}}, -\frac{1 - y}{y}, \frac{1}{1 - x}, \frac{1}{y \brk{1 - x y}}, 0}
        & \text{for DS1,}
        \\
        \brc{0, 0, 1, 1, 0, 0}
        & \text{for DS2.}
    \end{cases}
    \label{eq:RinvDS}
\end{align}
Very interestingly, the DS2 limit completely trivializes the $R$-invariants and makes the ratio function $\Enmhv\supbrk{1111}$ pure \brk{meaning it becomes
a $\Rationals$-linear sum of MPLs}. From this point on, we will focus on the DS2 limit as it greatly facilitates the search for the NMHV six-particle amplitude
within the DS function space $\funspace$.
Specifically, it allows us to use the same mixed form of eq. \eqref{eq:stfExp} for the collinear limit expansion of the entire amplitude, without having to take additional contributions from the rational factors into account.

\subsection{NMHV Wilson loop OPE}
\label{ssec:wlope}

Let us now proceed to discuss in detail how to determine the (1111) component of the six-particle NMHV amplitude in the DS2 limit with the help of the Wilson loop OPE. We start by briefly reviewing the latter approach, and especially the predictions it provides for the weak-coupling expansion of the amplitude, in integral form. We then proceed to trade them with infinite sum representations relying on Cauchy's residue theorem. Finally, we match them against the series representation of our ansatz, carried our in subsection~\ref{ssec:expansions}.

The Wilson loop OPE provides a non-perturbative description of the amplitude as a state sum expansion around the
collinear limit. The DS and collinear limits are closely related as can be seen from eqs. \eqref{eq:DS1def} and \eqref{eq:DS2def}. The DS
limit benefits from having only gluonic states with positive \brk{for DS1} or negative \brk{for DS2} helicities contributing to it. In the DS2 limit we will be focusing on, discussed in the previous subsection, the  $\brk{1111}$ component of the framed NMHV Wilson loop $\Wnmhv$ may be written as
\begin{align}
    \Wnmhv \xrightarrow[\text{DS2}]{}
    \sum_{N_- \ge 0} \> \Wnmhv_{N_-} =
    \sum_{N_- \ge 0} \sum_{L \ge 0} \> g^{2 L} \Wnmhv\supbrk{L}_{N_-},
    \label{eq:WnmhvDef}
\end{align}
where $\Wnmhv_{N_-}$ are contributions from multi-particle excitations consisting of $N_-$ negative helicity gluon bound states, further organized into $L$-loop contributions to their weak-coupling expansion. As we'll see below, $\Wnmhv_{N_-}$ starts contributing at $L = N_-^2 + N_-$ loops, and so for our purposes it will be sufficient to restrict to $N \le 2$, which provides an accurate desciption up to and including $L=11$ loops.

The one gluon bound state particle contribution $\Wnmhv_{1_-}$ is given by
\begin{align}
    \Wnmhv_{1_-} =
    \sum_{l_1 \ge 1}
    \brk{T F^{-1}}^{l_1}
    \int \frac{\dd u_1}{2 \pi}
    \>\>
    T^{\gamma_{l_1} \! \brk{u_1}}
    S^{\, \im \, p_{l_1} \! \brk{u_1}}
    \smtimes \mu_{l_1} \! \brk{u_1} \, h_{-l_1} \! \brk{u_1}
    ,
    \label{eq:w1def}
\end{align}
whereas the superposition of two gluon bound states $\Wnmhv_{2_-}$ by
\begin{align}
    \Wnmhv_{2_-} =
    \sum_{l_1 \ge l_2 \ge 1}
    \frac{1}{1 + \delta_{l_1, l_2}}
    \brk{T F^{-1}}^{l_1 + l_2}
    \int &\frac{\dd u_1 \> \dd u_2}{\brk{2 \pi}^2}
    \>\>
    T^{\gamma_{l_1} \! \brk{u_1} + \gamma_{l_2} \! \brk{u_2}}
    S^{\, \im \, p_{l_1} \! \brk{u_1} + \im \, p_{l_2} \! \brk{u_2}}
    \nn\\
    &\qquad\smtimes
    \frac{
        \mu_{l_1} \! \brk{u_1} \, \mu_{l_2} \! \brk{u_2} \, h_{-l_1} \! \brk{u_1} \, h_{-l_2} \! \brk{u_2}
    }{
        P_{l_1 | \, l_2}\brk{u_1 | u_2} \, P_{l_2 | \, l_1} \brk{u_2 | u_1}
    }.
    \label{eq:w2def}
\end{align}
In the above expressions, $p$ and $\gamma$ denote the momentum and quantum energy correction of the gluon excitations, whereas $\mu$, $P$ and $h$ are further physical quantities describing them which are known as the measure, pentagon transition and NMHV form factor. They have been derived at finite coupling in~\cite{Basso:2014nra}, and to leading quantum order, their perturbative expansions read
\begin{align}
    p_l \brk{u} &= 2 u + 2 \im g^2 \bigbrk{\psi\brk{\im \um{}} - \psi\brk{-\im \up{}}} + \order{g^4},
    \label{eq:pDef}
    \\
    \gamma_l \brk{u} &= 2 g^2 \bigbrk{\psi\brk{1 + \im \um{}} + \psi\brk{1 -\im \up{}} - 2 \psi\brk{1}} + \order{g^4},
    \label{eq:gammaDef}
    \\
    \mu_l \brk{u} &= \brk{-1}^l g^2 \> \bigbrk{\up{} \um{}}^{-2} \smtimes
    \frac{
        \Gfun\brk{1 + \im \um{}} \Gfun\brk{1 - \im \up{}}
    }{\Gfun\brk{l}} + \order{g^4},
    \label{eq:muDef}
    \\
    P_{l_1 | \, l_2} \brk{u_1 | u_2} &= \frac{\brk{-1}^{l_2}}{g^2} \>
    \up{1} \um{1} \up{2} \um{2} \smtimes
    \frac{\Gfun\brk{-\im \up{1} + \im \um{2}}}{\Gfun\brk{1 + \im \um{1}} \Gfun\brk{1 - \im \up{2}}} \smtimes
    \frac{\Gfun\brk{\im \um{1} - \im \um{2}}}{\Gfun\brk{1 - \im \up{1} + \im \up{2}}}
    + \order{g^4},
    \label{eq:Pdef}\\
 h_{l} \brk{u} &\defas \Bigbrk{
        \tfrac{1}{g^2} \, x\brk{\up{}} \, x\brk{\um{}}
    }^{\mathrm{sign}\brk{l}},
    \label{eq:formDef}
\end{align}
expressed in terms of the shifted rapidities
\be
u^\pm_k \defas u_k \pm \tfrac{\im \, l_k}{2}\,.
\ee
and Zhukowski variables
\be
x\brk{u} \defas \tfrac12 \bigbrk{u + \sqrt{u^2 - 4 g^2}} = u + \order{g^2}\,.
\ee

As a consequence, the $L$-loop contribution to the OPE integrals defined in eqs. \eqref{eq:w1def, eq:w2def} has the general structure
\begin{align}
    \Wnmhv\supbrk{L}_{1_-} =
    g^{2 L} \sum_{l_1 \ge 1}
    \Bigbrk{-\tfrac{T}{S F}}^{l_1}
    \int \frac{\dd u_1}{2 \pi}
    \>
    \frac{
        S^{2 \im \um{1}}
        \>
        \polyP_1\supbrk{L}
    }{
        \bigbrk{
            \up{1} \um{1}
        }^{2 L - 1}
    }
    \smtimes \frac{
        \Gfun\brk{1 - \im \up{1}} \Gfun\brk{1 + \im \um{1}}
    }{
        \Gfun\brk{l_1}
    }
    \label{eq:wlope1glu}
\end{align}
for one gluon bound state, and
\begin{align}
    &\Wnmhv\supbrk{L}_{2_-} =
    g^{2 L} \sum_{l_1 \ge l_2 \ge 1}
    \frac{1}{1 + \delta_{l_1, l_2}}
    \Bigbrk{-\tfrac{T}{S F}}^{l_1 + l_2}
    \int \frac{\dd u_1 \> \dd u_2}{\brk{2 \pi}^2}
    \>
    \frac{
        S^{2 \im \brk{\um{1} + \um{2}}}
        \>
        \polyP_2\supbrk{L}
    }{
        \bigbrk{
            \up{1} \um{1} \up{2} \um{2}
        }^{2 L - 7}
    }
    \smtimes
    \label{eq:wlope2glu}
    \\&\>
    \smtimes
    \frac{
        \Gfun\brk{1 - \im \up{1}}^2
        \Gfun\brk{1 + \im \um{1}}^2
        \Gfun\brk{1 - \im \up{2}}^2
        \Gfun\brk{1 + \im \um{2}}^2
    }{
        \Gfun\brk{l_1}
        \Gfun\brk{l_2}
        \Gfun\brk{-\im \up{1} + \im \um{2}}
        \Gfun\brk{\im \um{1} - \im \up{2}}
    }
    \cdot
    \frac{
        \Gfun\brk{1 + \im \up{1} - \im \up{2}}
        \Gfun\brk{1 - \im \up{1} + \im \up{2}}
    }{
        \Gfun\brk{\im \um{1} - \im \um{2}}
        \Gfun\brk{- \im \um{1} + \im \um{2}}
    }
    \nn
\end{align}
for two excitations, which becomes non-zero starting from the $L = 6$ loop order. Here the $\polyP_{k}\supbrk{L}$ symbols
collect factors with polynomial dependence on the shifted rapidities $u^{\pm}_{k}$, the logarithms of kinematical variables $\brc{\log\brk{T},
\log\brk{S}}$, and on polygamma functions $\brc{\psi^{\bigcdot}\brk{\mp u^\pm_k}, \psi^{\bigcdot}\brk{1 \mp u^\pm_k}}$, which appear at higher order in the perturbative expansions~\eqref{eq:pDef}-\eqref{eq:formDef}.
For example, the first two non-zero values for the $\polyP\supbrk{L}_1$ read:
\begin{align}
    \polyP_1\supbrk{2} &= 1,
    \\
    \polyP_1\supbrk{3} &=
    \tfrac72 \brk{\um{1}}^2 + \um{1} \up{1} + \tfrac72 \brk{\up{1}}^2
    - \brk{\um{1} \up{1}}^2 \Bigbrk{
        -\tfrac{\pi^2}{3} + \bigbrk{
            \psi\supbrk{1}\brk{1 + \im \um{1}}
            + \psi\supbrk{1}\brk{1 - \im \up{1}}
        }
        \nn\\&\qquad
        + \tfrac12 \bigbrk{
            \psi\brk{1 + \im \um{1}}
            + \psi\brk{1 - \im \up{1}}
        }^2
        + \tfrac12 \bigbrk{
            \psi\brk{\im \um{1}}
            - \psi\brk{-\im \up{1}}
        }^2
        \nn\\&\qquad
        + 2 \bigbrk{
            \psi\brk{1 + \im \um{1}}
            + \psi\brk{1 - \im \up{1}}
        } \log\brk{T}
        + 2 \bigbrk{
            \psi\brk{\im \um{1}}
            - \psi\brk{-\im \up{1}}
        } \log\brk{S}
    }
    \nn\\&\quad
    + \EulerGamma \> \brk{\ldots},
\end{align}
and similarly for $\polyP\supbrk{L}_2$:
\begin{align}
    \polyP_2\supbrk{6} &= 1,
    \\
    \polyP_2\supbrk{7} &= \tfrac{11}{2} \Bigsbrk{
        \brk{\um{1} \up{1} \um{2}}^2
        + \brk{\um{1} \up{1} \up{2}}^2
        + \brk{1 \leftrightarrow 2}
    }
    \nn\\&\quad
    + \um{1} \up{1} \um{2} \up{2} \bigbrk{
        \um{1} \up{1}
        + \um{1} \um{2}
        + \um{1} \up{2}
        + \up{1} \um{2}
        + \up{1} \up{2}
        + \um{2} \up{2}
    }
    \nn\\&\quad
    - \brk{\um{1} \up{1} \um{2} \up{2}}^2 \Bigbrk{
        - \tfrac{4 \pi^2}{3}
        + \Bigsbrk{
            \tfrac12 \bigbrk{
                \psi\brk{1 + \im \um{1}}
                + \psi\brk{1 - \im \up{1}}
            }^2
            + \tfrac12 \bigbrk{
                \psi\brk{\im \um{1}}
                - \psi\brk{-\im \up{1}}
            }^2
            \nn\\&\qquad
            +2 \bigbrk{
                \psi\supbrk{1}\brk{1 + \im \um{1}}
                + \psi\supbrk{1}\brk{1 - \im \up{1}}
            }
            -2 \psi\brk{\im \um{1}} \psi\brk{-\im \up{1}}
            \nn\\&\qquad
            +2 \bigbrk{
                \psi\brk{1 + \im \um{1}}
                + \psi\brk{1 - \im \up{1}}
            } \log\brk{T}
            +2 \bigbrk{
                \psi\brk{\im \um{1}}
                - \psi\brk{-\im \up{1}}
            } \log\brk{S}
            + \brk{1 \leftrightarrow 2}
        }
        \nn\\&\qquad
        - \bigbrk{\psi\brk{1 + \im \um{1}} + \psi\brk{1 - \im \up{1}}} \bigbrk{\psi\brk{1 + \im \um{2}} + \psi\brk{1 - \im \up{2}}}
        \nn\\&\qquad
        - \bigbrk{\psi\brk{\im \um{1}} - \psi\brk{-\im \up{1}}} \bigbrk{\psi\brk{\im \um{2}} + \psi\brk{- \im \up{2}}}
        }
        \nn\\&\quad
        + \EulerGamma \> \brk{\ldots}.
\end{align}
Here the ellipsis $ \> \brk{\ldots}$ hide terms proportional to the Euler-Mascheroni constant $\EulerGamma$, which always drop out
from the final answer.
Note that the last fraction in eq. \eqref{eq:wlope2glu} can actually be simplified to just a polynomial \cite{Drummond:2015jea}:
\begin{align}
    \frac{
        \Gfun\brk{1 + \im \up{1} - \im \up{2}}
        \Gfun\brk{1 - \im \up{1} + \im \up{2}}
    }{
        \Gfun\brk{\im \um{1} - \im \um{2}}
        \Gfun\brk{- \im \um{1} + \im \um{2}}
    }
    =
    \brk{-1}^{l_1 - l_2 + 1} \brk{\im \up{1} - \im \up{2}} \brk{\im \um{1} - \im \um{2}},
\end{align}
which means that it does not contribute any additional poles to the integrand.

After the perturbative expansion of eqs. \eqref{eq:wlope1glu} and \eqref{eq:wlope2glu} is achieved to the desired loop order $L$, our next step
is application of the Cauchy's residue theorem: for $S > 1$ the contour of $u_k$-integration is closed in the upper
half-plane and the residues at poles $u_k = \im \brk{j_k + \tfrac{l_k}{2}}$ are collected. Following the ideas of \cite{Papathanasiou:2013uoa,
Drummond:2015jea}, we first change the variables in the integrands according to:
\begin{align}
    u_k = \eps_k + \im \brk{j_k + \tfrac{l_k}{2}}, \quad
    u^\pm_k = \eps_k + \im j_k +
    \begin{cases}
        \im \, l_k & \text{for $+$}\\
        0     & \text{for $-$}
    \end{cases} \quad
    \text{for} \quad
    j_k \in \Z_{\ge 0}.
    \label{eq:resDef}
\end{align}
Next we expand the integrands in the $\eps_k \to 0$ limit and pick the coefficient of the $\eps_k^{-1}$ term, which gives an analytic expression for
the residues as functions on an integer lattice of the summation variables $\brc{j_1, l_1}$ in the one excitation case, and $\brc{j_1, j_2, l_1, l_2}$
in the case of two excitations with $j_k \ge 0$ and $l_1 \ge l_2 \ge 1$.

\begin{figure}[t]
    \centering
    \includegraphics[align=c]{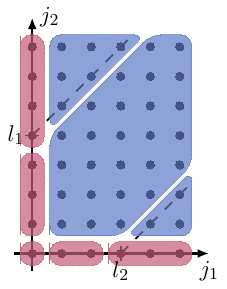}
    \caption{
        A graphical depiction of the eight regions of the $\Wnmhv_{2_-}$ summation variables presented on the $\brc{j_1, j_2}$-plane, parametrized by the helicities
        $\brc{l_1, l_2}$ and listed in \protect\tabref{tab:sign}.
    }
    \label{fig:regions}
\end{figure}

Special care is needed for the $j_k = 0$ boundary, where the corresponding shifted rapidity $\um{k}$ in the denominators of
the integrands of eqs. \eqref{eq:wlope1glu} and \eqref{eq:wlope2glu} develop additional $\eps_k$ singularities.
The lattice of summation variables $\brc{j_1, l_1}$ gets split into 2 regions with respect to the
$j_1 > 0$ condition: one region is $\brc{j_1 = 0}$ and the other is $\brc{j_1 > 0}$.
The first two $u$-dependent $\Gfun$-functions in the denominator of eq. \eqref{eq:wlope2glu} also require separate treatment,
as they reduce the overall $\eps_k$-divergence when $-\im \up{1} + \im \um{2} \in \Integers_{\ge 0}$ or $\im \um{1} - \im \up{2} \in \Integers_{\ge 0}$.
When combined, these conditions split the lattice of summation variables $\brc{j_1, j_2, l_1, l_2}$ into 8 regions $\brc{\mathrm{I}, \ldots,
\mathrm{VIII}}$, as illustrated in \figref{fig:regions}, which we distinguish by the values of the following 4-tuple of predicates:
\begin{align}
    \brc{j_1 > 0, j_2 > 0, j_2 < j_1 + l_1, j_2 > j_1 - l_2}.
    \label{eq:sign}
\end{align}
In \tabref{tab:sign} we list the eight possible combinations of values of these predicates as well as the total number of the lattice points that
satisfy them truncated at $j_1 + j_2 + l_1 + l_2 \le \Nopt$, where we took the optimal expansion order $\Nopt = 25$ from
\tabref{tab:order}.

\begin{table}
    \centering
    \begin{tabular}{c|cccc|c}
        \toprule
        region  & $j_1 > 0$ & $j_2 > 0$ & $j_2 < j_1 + l_1$ & $j_2 > j_1 - l_2$ & number of points \\
        \midrule
        {\color{blue1!90!black} I}      & \cmark   & \cmark   & \cmark   & \cmark   & $3360$ \\
        {\color{blue1!90!black} II}     & \cmark   & \cmark   & \cmark   & \xmark   & $2272$ \\
        {\color{blue1!90!black} III}    & \cmark   & \cmark   & \xmark   & \cmark   & $1232$ \\
        {\color{magenta1} IV}           & \cmark   & \xmark   & \cmark   & \cmark   & $358$ \\
        {\color{magenta1} V}            & \cmark   & \xmark   & \cmark   & \xmark   & $864$ \\
        {\color{magenta1} VI}           & \xmark   & \cmark   & \cmark   & \cmark   & $764$ \\
        {\color{magenta1} VII}          & \xmark   & \cmark   & \xmark   & \cmark   & $458$ \\
        {\color{magenta1} VIII}         & \xmark   & \xmark   & \cmark   & \cmark   & $156$ \\
        \bottomrule
    \end{tabular}
    \caption{
        The eight regions of the summation variables $\brc{j_1, j_2, l_1, l_2}$ and the values of the predicates
        introduced in eq.  \protect\eqref{eq:sign} that separate them.
        Here we denote $\brc{\text{\cmark}, \text{\xmark}} = \brc{\text{true}, \text{false}}$.
        The total number of points \brk{or terms in the sums} in each region is evaluated for the expansion
        order $x^{25}$, which corresponds to $j_1 + j_2 + l_1 + l_2 \le N_\text{opt}(12)=25$. In \protect\figref{fig:regions} we also illustrate these
        regions on the $\brc{j_1, j_2}$-plane.
    }
    \label{tab:sign}
\end{table}

Finally we arrive at the sum over residues for the one gluon OPE integral shown in eq. \eqref{eq:wlope1glu} containing the following two terms:
\begin{align}
    \Wnmhv\supbrk{L}_{1_-} =
    \sum_{j_1, l_1 \ge 1}
    \> x^{j_1 + l_1} \brk{1 - y}^{j_1}
    \smtimes \frac{
        \polyQ\supbrk{L}_{1}
    }{
        \bigbrk{j_1 \brk{j_1 + l_1}}^{2 L - 1}
    }
    \cdot \frac{
        \Gfun\brk{j_1 + l_1}
    }{
        \Gfun\brk{l_1}
        \Gfun\brk{1 + j_1}
    }
    +
    \sum_{l_1 \ge 1}
    \> x^{l_1}
    \smtimes \frac{
        \polyQ\supbrk{L}_{2}
    }{
        l_1^{2 L + 2}
    }
    \label{eq:w1sum}
\end{align}
where we collected all the dependence on $\brc{j_1, l_1}$ parameters, as well as $\brc{\psi^{\bigcdot}\brk{j_1}, \psi^{\bigcdot}\brk{j_1 + l_1}}$
and the $\brc{\log\brk{T}, \log\brk{S}}$ logarithms in two factors $\polyQ\supbrk{L}_1$ for the $j_1 > 0$ region and $\polyQ\supbrk{L}_2$ for the $j_1 = 0$ region.
Similarly, the residue sum for the two gluon OPE integral of eq. \eqref{eq:wlope2glu} involves eight terms, one for each region shown in \tabref{tab:sign}:
\begin{align}
    &\Wnmhv\supbrk{L}_{2_-} =
    \sum_{\brc{j_1, j_2, l_1, l_2} \in \mathrm{I}}
    \> x^{j_1 + j_2 + l_1 + l_2} \brk{1 - y}^{j_1 + j_2}
    \smtimes \frac{
        \polyR\supbrk{L}_{\mathrm{I}}
    }{
        \bigbrk{j_1 j_2 \brk{j_1 + l_1} \brk{j_2 + l_2}}^{2 L - 6}
    }
    \cdot
    \label{eq:w2sum}
    \\&\>
    \cdot \frac{
        \Gfun\brk{j_1 + l_1}^2 \Gfun\brk{j_2 + l_2}^2
    }{
        \Gfun\brk{l_1}
        \Gfun\brk{l_2}
        \Gfun\brk{1 + j_1}^2
        \Gfun\brk{1 + j_2}^2
        \Gfun\brk{j_1 - j_2 + l_1}
        \Gfun\brk{-j_1 + j_2 + l_2}
    }
    + \text{other regions $\brc{\mathrm{II}, \ldots, \mathrm{VIII}}$},
    \nn
\end{align}
where we hide all the complicated dependence on the summation variables $\brc{j_1, j_2, l_1, l_2}$ and the polygamma functions
$\brc{
    \psi^{\bigcdot}\brk{j_k},
    \psi^{\bigcdot}\brk{j_k + l_k},
    \psi^{\bigcdot}\brk{j_1 - j_2 + l_1},
    \psi^{\bigcdot}\brk{-j_1 + j_2 + l_2}
}$
in $\brc{\polyR\supbrk{L}_{\mathrm{I}}, \ldots, \polyR\supbrk{L}_{\mathrm{VIII}}}$ factors.

While in~\cite{Papathanasiou:2013uoa, Drummond:2015jea} it was shown that the sum representation of the one-gluon contribution \eqref{eq:w1sum} can be evaluated in terms of MPLs with the help of the algorithms of ref.\cite{Moch:2001zr}, to our knowledge there exists no direct method for similarly evaluating the two-gluon contribution, eq. \eqref{eq:w2def}. It is for this reason that we will instead choose to resort to the bootstrap method.

In summary, by virtue of eq.~\eqref{eq:WnmhvDef} the above residue sum representations of the OPE
gluonic contributions $\Wnmhv\supbrk{L}_1$ and $\Wnmhv\supbrk{L}_2$ yield the (1111) component of the NMHV Wilson loop $\Wnmhv$ in the DS2 limit as a series expansion around the collinear limit. Converting the latter to the BDS-like normalization according to eq.~\eqref{eq:EmhvDef}, and recalling that it is a pure function as a consequence of eq.~\eqref{eq:RinvDS}, we then equate it to an ansatz built out of the DS functions we constructed in \secref{sec:bootstrap}, in their series expansion representation~\eqref{eq:stfExp}. This fixes all coefficients of our ansatz, and using the MPL representation of our DS functions, we thus determine the (1111) component of $\Enmhv\supbrk{L}$ in this form.

Before presenting our results in the next section, let us also make some further comments on our computational setup. The form of the sums of eqs. \eqref{eq:w1sum} and \eqref{eq:w2sum} indicates that in every term the power of the $x$-variable is bigger than the
power of the $y$-variable. In our ansatz, they will thus correspond to the $\ndiagSt$ ``non-diagonal'' part of the expansions of eq. \eqref{eq:stfExp}, implying that its ``diagonal'' part is zero. This makes our ansatz expansion of the form~\eqref{eq:stfExp} well-suited for matching onto eqs. \eqref{eq:w1sum} and \eqref{eq:w2sum}, and further implies that we may bound the summation ranges of the latter according to $j_1 + l_1 \le \Nopt$ and $j_1 + j_2 + l_1 + l_2 \le \Nopt$, where $\Nopt$ is the optimal expansion order discussed in subsection \ref{ssec:expansions} and shown in \tabref{tab:order}.

\section{Results}
\label{sec:results}

In this work we constructed a space of functions $\funspace$ relevant for the DS limit of the six gluon scattering amplitude in $\superN = 4$
super-Yang-Mills theory through transcendental weight $\weight = 12$.
We applied the $\funspace$ space to \brk{partial} calculation of the NMHV component of the six gluon amplitude up to $L \le 8$ loops.

In \secref{ssec:coprods} we employed extended Steinmann and analyticity conditions in order to construct the coproduct representation of the
$\funspace$ space.  In \secref{ssec:funs} and \secref{ssec:expansions} we present two closely related realizations of the $\funspace$: one is in terms
of multiple polylogarithms, and one in terms of power-and-log expansions around the $\brc{u, v, w} \to \brc{0, 1, 0}$ kinematical point.  The latter
is then combined with the residue sum representation of the one and two gluon contributions to the Wilson Loop Operator Product Expansion, which is
explained in some detail in \secref{sec:hex}. Combination of these two techniques
allows us to determine the six gluon NMHV amplitude in the DS limit to high loop order,
find new \brk{potentially all-loop} patterns in the analytical form of the amplitude in the $\brc{u, v, w} \to 0$ origin limit,
and pinpoint the excessive elements of the $\funspace$ that should be irrelevant for amplitude calculations
as will be explained next.

\subsection{The NMHV amplitude up to 8 loops and weight 12}
As the main application of our DS space of functions $\funspace$, we have computed the (1111) component of the NMHV superamplitude $\Enmhv\supbrk{L}$ in the double scaling limit $w \to 0$, more precisely one of its two parity images defined in eq.~\eqref{eq:DS2def}. This quantity has the following natural decomposition,
\begin{align}
    \Enmhv\supbrk{L}\brk{u, v, w} \xrightarrow[]{(1111)}
    \sum_{\weight = 0}^{2 L}
    \Enmhv\supbrk{L}_\weight\brk{u, v} \cdot
    \bigbrk{\log\brk{w}}^{2 L - \weight},
\end{align}
where each coefficient $\Enmhv\supbrk{L}_{\weight}\brk{u, v}$ of a large logarithm $\log\brk{w}^{2 L - p}$ is a pure transcendental function of weight
$\weight$. In particular, we have determined $\Enmhv\supbrk{L}_{\weight}$ for any $L\le8$ and $p\le 12$, namely the full component for $L\le 6$, as well as all terms with two (four) or more powers of $\log(w)$ at $L=7$ ($L=8$). The $p\le 12$ restriction is due to the maximal weight we have explicit MPL representations for $\funspace$ so far, whereas the  $L\le 8$ restriction is due to the size of the expressions predicted by the Wilson loop OPE. Nevertheless, we are hopeful that the new results we have obtained will offer valuable boundary data and consistency checks for NMHV hexagon in general kinematics at seven~\cite{DDunpublished} and eight loops. We provide expressions
$\sum_{L = 0}^6 \sum_{\weight = 0}^{2 L} g^{2 L} \Enmhv\supbrk{L}_{\weight} \cdot \bigbrk{\log\brk{w}}^{2 L - \weight}$ and
$\sum_{L = 7}^8 \sum_{\weight = 0}^{12}  g^{2 L} \Enmhv\supbrk{L}_{\weight} \cdot \bigbrk{\log\brk{w}}^{2 L - \weight}$
for the full $\brk{1111}$ component of the NMHV amplitude in the DS limit up to 6 loops and partial results at 7 and 8 loops in the files
\ancillary{EDS-L1-8-w0-12.m}.

Let us also describe the checks we have performed on our answers. First off, we confirm BDS-normalized component vanishes in the {\color{green1} soft} limit  boundary $x \to 0$ limit or, equivalently,
\begin{align}
    \Enmhv\supbrk{L} \xrightarrow[\text{{\color{green1} soft} \& DS2}]{} \Bigbrk{e^{\tfrac14 \gcusp \Emhv\supbrk{1}}}\supbrk{L},
    \label{eq:EnmhvResults}
\end{align}
as expected. Furthermore, we find perfect agreement with existing results on the collinear limit expansion of the NMHV amplitude, namely the $\brc{T, T^2, T^3}$ terms for $L \le 4$ from \filename{eEEtT3.m} of \cite{Caron-Huot:2016owq}, as well as the first two $\brc{T, T^2}$ terms for $L \le 6$ from \filename{W1111L0-6.m} of \cite{Caron-Huot:2019vjl} (this time for the individual OPE contributions $\Wnmhv\supbrk{L}_{1_-}$ and
$\Wnmhv\supbrk{L}_{2-}$). Additional checks were also performed in an interesting boundary point of the DS limit, that we discuss next.

\subsection{The origin limit}
The origin limit $\brc{u, v, w} \to 0$ was first analyzed in~\cite{Caron-Huot:2019vjl}, where it was observed that at weak coupling the MHV six-particle amplitude takes a very simple (Sudakov-like) form of exponentiated double logarithms,
\begin{align}
    \log\brk{\Emhv} = -\tfrac{\Gamma_{\text{oct}}}{24} l^2_{u v w} - \tfrac{\Gamma_{\text{hex}}}{24} \bigbrk{
        l^2_{u / v} + l^2_{v / w} + l^2_{w / u}
    } + C_0 + \order{u},
\end{align}
in terms of certain quantities $\brc{\Gamma_{\text{oct}}, \Gamma_{\text{hex}}, C_0}$ that only depend on the coupling. More recently these quantities have been conjectured at finite coupling~\cite{Basso:2020xts}, based on a Wilson loop OPE resummation procedure morally similar to the one considered in this paper, and backed by further strong coupling analysis via the gauge/string duality.

Although in the latter paper it was also pointed out that the NMHV amplitude no longer displays this exponentiation of double logarithms, it may still be interesting to look for patterns that may hint to an all-loop description. Clearly, the origin limit is a boundary of the DS limit, corresponding to $y\to 1^+$, $x\to -\infty$ in the variables~\eqref{eq:xytou} underlying the construction of our DS function space. We have indeed taken this limit of the general result discussed in the previous section, and through four loops it reads
\begin{align}
    &\Enmhv \xrightarrow[\text{DS2 \& origin}]{(1111)}
    1 + g^2 \Bigsbrk{-2 \zeta_2+l_u l_v-\frac{l_u^2}{2}-\frac{l_v^2}{2}-\frac{l_w^2}{2}}
    \nn\\
    &+ g^4 \Bigsbrk{
        \frac{l_w^4}{8}
        +l_w^2 \brk{2 \zeta_2-\frac{l_u l_v}{2}+\frac{l_u^2}{4}+\frac{l_v^2}{4}}
        \nn\\&\quad
        -\frac{1}{2} l_u^3 l_v+\frac{1}{2} l_u^2 l_v^2-\frac{1}{2} l_u l_v^3+\frac{l_u^4}{8}+\frac{3}{2} \zeta_2 l_v^2+\zeta_3 l_v+\frac{l_v^4}{8}
        +\frac{39 \zeta_2^2}{10}+\frac{3}{2} \zeta_2 l_u^2+\zeta_3 l_u-5 \zeta_2 l_u l_v
    }
    \nn\\
    &+ g^6 \Bigsbrk{
        -\frac{l_w^6}{48}
        +l_w^4 \brk{-\frac{3 \zeta_2}{4}+\frac{l_u l_v}{8}-\frac{l_u^2}{16}-\frac{l_v^2}{16}}
        \nn\\&\quad
        +l_w^2 \brk{
            -\frac{167 \zeta_2^2}{20}-\frac{5}{4} \zeta_2 l_u^2-\frac{\zeta_3 l_u}{2}+\frac{7}{2} \zeta_2 l_u l_v+\frac{1}{4} l_u^3 l_v
            -\frac{1}{4} l_u^2 l_v^2+\frac{1}{4} l_u l_v^3-\frac{l_u^4}{16}-\frac{5}{4} \zeta_2 l_v^2-\frac{\zeta_3 l_v}{2}-\frac{l_v^4}{16}
        }
        \nn\\&\quad
        +l_w \brk{
            \frac{1}{2} \zeta_3 l_u^2+2 \zeta_2^2 l_u+\frac{1}{2} \zeta_2 l_u^2 l_v+\frac{1}{2} \zeta_2 l_u l_v^2+2 \zeta_2^2 l_v
            +\frac{1}{2} \zeta_3 l_v^2
        }
        \nn\\&\quad
        +\frac{1}{8} l_u^5 l_v-\frac{3}{16} l_u^4 l_v^2+\frac{5}{18} l_u^3 l_v^3-\frac{3}{16} l_u^2 l_v^4+\frac{1}{8} l_u l_v^5
        -\frac{l_u^6}{48}-\frac{69}{20} \zeta_2^2 l_v^2-\frac{1}{2} \zeta_2 l_v^4-\zeta_3 l_v^3-6 \zeta_2 \zeta_3 l_v-8 \zeta_5 l_v
        -\frac{l_v^6}{48}
        \nn\\&\quad
        -\frac{527 \zeta_2^3}{105}-\frac{1}{2} \zeta_2 l_u^4-\zeta_3 l_u^3-\frac{69}{20} \zeta_2^2 l_u^2-6 \zeta_2 \zeta_3 l_u
        -8 \zeta_5 l_u+\frac{19}{6} \zeta_2 l_u^3 l_v-\frac{5}{2} \zeta_2 l_u^2 l_v^2+\frac{271}{10} \zeta_2^2 l_u l_v
        +\frac{19}{6} \zeta_2 l_u l_v^3
    }
    \nn\\
    &+ \order{g^8},
    \label{eq:EnmhvOrigin}
\end{align}
where the symbol $l_x$ represents a simple logarithm: $l_x \defas \log\brk{x}$ as defined in eq. \eqref{eq:notation}. Higher loop order $L \le 8$
corrections to eq. \eqref{eq:EnmhvOrigin} bounded by transcendental weight $\weight \le 12$ can be found in the ancillary file
\ancillary{EDS-origin-L1-8-w0-12.m}.

In this manner, we confirm that no additional cancellations occur for the (1111) component, that would lead to a Sudakov-like form similar to the MHV amplitude at the origin. We have also further vetted our result by comparing it with the origin limit of the NMHV amplitude in general kinematics through six loops~\cite{LDprivate,DDunpublished}, finding perfect agreement.

While it may be worthwhile to also look at other components for additional cancellations, already from the current data we observe an interesting general pattern for the three highest powers of divergent logarithms for the NMHV/MHV ratio function. In particular, up to $L \le 6$ loops we find\footnote{We acknowledge discussions with Lance Dixon, who first noticed the leading $1/(L!)^2$ behavior on the $u=v=w$ line of the NMHV amplitude at the origin, motivating us to carry out the following analysis.}
\begin{align}
    &\Enmhv / \Emhv \xrightarrow[\text{DS2 \& origin}]{(1111)}
    \sum_{L = 0}^{6} g^{2 L} \Bigsbrk{
        \frac{\brk{l_u l_v}^{L}}{\brk{L!}^2}
        \nn\\
        &\quad + \zeta_2 \Bigbrk{
            - \frac{\brk{l_u l_v}^{L - 2} \brk{l_u + l_v} l_w}{\brk{L - 2}! \brk{L - 1}!}
            - \frac{\brk{l_u l_v}^{L - 2} \brk{\brk{L - 1}^2 l_u^2 + \brk{4 \brk{L - 1}^2 + 3 \brk{L - 1} - 1} l_u l_v + \brk{L - 1}^2 l_v^2}}{\brk{L - 1}! L!}
        }
        \nn\\
        &\quad + \zeta_3 \Bigbrk{
            \frac{\brk{l_u l_v}^{L - 3} \brk{l_u^2 + l_v^2} l_w}{\brk{L - 2}! L!}
            - \frac{
                \brk{l_u l_v}^{L - 3} \brk{l_u + l_v} \brk{\brk{L - 2} l_u^2 - \brk{L - 1} l_u l_v + \brk{L - 2} l_v^2}
            }{
                \brk{L - 2}! \brk{L - 1}!
            }
        }
        \nn\\
        &\quad + \order{\brk{l_u l_v}^{2 L - 4}}
    }
    \nn\\
    &\> + \order{g^{14}},
\end{align}
It may thus well be that this pattern persists at higher loops, especially given that the two-gluon OPE excitation already contributes at $L=6$ loops in the DS limit.

\subsection{Further refinements of the $\funspace$ space}
\label{ssec:saturation}
In this final subsection, let us also discuss how it may be possible to further reduce the size of the $\funspace$ functional space, so as to make bootstrapping at higher loops more tractable. Our
construction of $\funspace$ includes the entire basis of independent MZVs from eq. \eqref{eq:mzvbasis}, while we notice only the ordinary zeta values appear in the
NMHV remainder $\Enmhv$ function. This is an indication that perhaps only the latter are needed as independent constants in our space.

To answer such questions about the potential redundancy of our space,  we studied the ``nested derivatives'' or
the $\brc{n, 1, \ldots, 1}$ coproduct components of the NMHV amplitude expressed in terms of the elements of the $\funspace$ space, following the blueprint of \cite{Caron-Huot:2019bsq}.  In
\tabref{tab:saturation} we report our findings on the dimensions of the minimal subspaces, needed to match the $\Enmhv\supbrk{6}_\weight$ and its coproducts.
Note how these dimensions saturate for increasing weight $\weight$ and fixed $n^{\text{th}}$ coproduct component for
\begin{align}
    p = 2 n + 2
\end{align}
at the latest, and the saturated value is smaller than the dimension of the corresponding DS subspace $\dim_{\Rationals} \funspace_n$, which implies
existance of unnecessary elements in the $\funspace$ space.
We find that the first such ``extra'' function $f^{\brk{3}, \text{extra}}_1 \in \funspace_3$ at weight $\weight = 3$ that does not appear in the
NMHV amplitude reads:
\begin{align}
    f^{\brk{3}, \text{extra}}_1 = \stf\supbrk{3}_2 \equiv \stf\supbrk{1}_1 \zeta_2 = l_{1 - x y} \zeta_2,
\end{align}
where $\yinv \defas \tfrac1y$, $\Gx_{\vec{X}} \defas G\brk{\vec{X}; x}$, and $l_x \defas \log\brk{x}$. Similarly, at weight $\weight = 4$ the space of ``extra'' functions is spanned by the following four elements:
\begin{align}
    f^{\brk{4}, \text{extra}}_1 &= \stf\supbrk{1}_1 \zeta_3 =
    l_{1 - x y} \> \zeta_3,
    \nn\\
    f^{\brk{4}, \text{extra}}_2 &= \stf\supbrk{2}_1 \zeta_2 =
    \Gx_{0 \yinv} \> \zeta_2,
    \nn\\
    f^{\brk{4}, \text{extra}}_3 &= \stf\supbrk{2}_2 \zeta_2 =
    \bigbrk{\Gx_{0 \yinv}-\Gx_{1 \yinv}-\Gx_{0 1}+l_{1-x} l_{x (1-y)}} \> \zeta_2,
    \nn\\
    f^{\brk{4}, \text{extra}}_4 &= \stf\supbrk{2}_4 \zeta_2 =
    \bigbrk{l_{x (1-y)} l_{1-x y}-l_{1-x y}^2} \> \zeta_2,
\end{align}
and at weight $\weight = 5$ we found the following eleven functions:
\begin{align}
    \brc{f^{\brk{5}, \text{extra}}_1, \ldots, f^{\brk{5}, \text{extra}}_{11}} =
    \brc{
        \stf\supbrk{3}_1 \zeta_2,
        \stf\supbrk{2}_1 \zeta_3,
        \stf\supbrk{3}_4 \zeta_2,
        \stf\supbrk{3}_5 \zeta_2,
        \stf\supbrk{2}_2 \zeta_3,
        \stf\supbrk{3}_9 \zeta_2,
        \nn\\
        \stf\supbrk{3}_7 \zeta_2,
        \stf\supbrk{3}_8 \zeta_2,
        \stf\supbrk{3}_{10} \zeta_2
        \stf\supbrk{2}_4 \zeta_3,
        \stf\supbrk{3}_{11} \zeta_2
    }
\end{align}
Interestingly, we see that while $\zeta_2$ and $\zeta_3$ are needed as independent constants in our space, not all of their products with non-constant DS functions appear at higher weight. From these findings we conclude that our DS functional space $\funspace$ is indeed overcomplete, at least for the problem of bootstrapping the NMHV remainder function in
$\superN = 4$ SYM. It would be interesting to perform a more careful study of the intricate interplay between the integrability and extended Steinmann relations of eqs.
\eqref{eq:integrability} and \eqref{eq:steinmann}, the branch cut conditions of eq. \eqref{eq:brcond}, and the coaction principle
\cite{Caron-Huot:2019bsq} in order to further perfect the DS bootstrap. We hope that it will provide valuable insights useful for the
general hexagon bootstrap program as well as for other computational problems in high energy physics.

\begin{table}
    \centering
    \begin{tabular}{l|cccccc ccccc}
        \toprule
        $n^{\text{th}}$ comp.      & 1 & 2 & 3 & 4 & 5 & 6 & 7 & 8 & 9 & 10 & 11 \\
        \midrule
        $\weight = 2$   & 1 &   &   &   &   &   &   &   &   &    &    \\
        $\weight = 3$   & 1 & 1 &   &   &   &   &   &   &   &    &    \\
        $\weight = 4$   & 2 & 4 & 2 &   &   &   &   &   &   &    &    \\
        $\weight = 5$   & 2 & 4 & 5 & 2 &   &   &   &   &   &    &    \\
        $\weight = 6$   & 2 & 5 & 9 & 5 & 2 &   &   &   &   &    &    \\
        $\weight = 7$   & 2 & 5 & 10& 11& 5 & 2 &   &   &   &    &    \\
        $\weight = 8$   & 2 & 5 & 11& 19& 12& 5 & 2 &   &   &    &    \\
        $\weight = 9$   & 2 & 5 & 11& 22& 28& 12& 5 & 2 &   &    &    \\
        $\weight = 10$  & 2 & 5 & 11& 22& 40& 28& 12& 5 & 2 &    &    \\
        $\weight = 11$  & 2 & 5 & 11& 22& 44& 60& 28& 12& 5 & 2  &    \\
        $\weight = 12$  & 2 & 5 & 11& 22& 45& 79& 60& 28& 12& 5  & 2  \\
        \midrule
        functions       & 2 & 5 & 12& 26& 56&116&236&474&943&1867&3686 \\
        \bottomrule
    \end{tabular}
    \caption{
        The number of independent $\brc{n, 1, \ldots, 1}$ coproduct components of all $\log{w}$ coefficients $\Enmhv\supbrk{6}_{\weight}$ of the
        $L = 6$ loop NMHV amplitude from eq. \protect\eqref{eq:EnmhvResults}. For the reader's convenience we copied the total number of
        DS functions $\stf\supbrk{\weight}$ at each weight $\weight$ from \tabref{tab:coprodsSize}.
    }
    \label{tab:saturation}
\end{table}

\section*{Acknowledgements}
GP would like to thank Benjamin Basso, Simon Caron-Huot, Lance Dixon, Falko Dulat, Matt von Hippel and Andrew McLeod for collaboration on closely
related topics, as well as Lance Dixon and \"Omer G\"urdo\u gan for discussions on related $A_2$ function counts. VC thanks Sven-Olaf Moch for helpfull discussions and encouragement.
The authors acknowledge support from the Deutsche Forschungsgemeinschaft \brk{DFG} under Germany’s Excellence Strategy EXC 2121 ``Quantum
Universe'' 390833306.
The work of VC has also been supported by the DFG under grant number MO 1801/2-1.

\appendix

\section{The Multiple Zeta Value basis}
\label{app:mzvs}
In this work we used the basis and reduction rules for MZVs from the MZV datamine project \cite{Blumlein:2009cf}. The basis reads as follows:
\begin{align}
    \bigbrc{
        \brc{\zeta_{2}},
        \brc{\zeta_{3}},
        \brc{\zeta_{2}^2},
        \brc{\zeta_{2}\zeta_{3}, \zeta_{5}},
        \brc{\zeta_{2}^3, \zeta_{3}^2},
        \brc{\zeta_{2}^2\zeta_{3}, \zeta_{2}\zeta_{5}, \zeta_{7}},
        \brc{\zeta_{2}^4, \zeta_{2}\zeta_{3}^2, \zeta_{3}\zeta_{5}, \zeta_{5,3}}, \nn\\
        \brc{\zeta_{2}^3\zeta_{3}, \zeta_{3}^3, \zeta_{2}^2\zeta_{5}, \zeta_{2}\zeta_{7}, \zeta_{9}},
        \brc{\zeta_{2}^5, \zeta_{2}^2\zeta_{3}^2, \zeta_{2}\zeta_{3}\zeta_{5}, \zeta_{5}^2, \zeta_{3}\zeta_{7}, \zeta_{2}\zeta_{5,3}, \zeta_{7,3}}, \nn\\
        \brc{\zeta_{11}, \zeta_{2}^4\zeta_{3}, \zeta_{2}\zeta_{3}^3, \zeta_{2}^3\zeta_{5}, \zeta_{3}^2\zeta_{5}, \zeta_{3}\zeta_{5,3}, \zeta_{5,3,3}, \zeta_{2}^2\zeta_{7}, \zeta_{2}\zeta_{9}}, \nn\\
        \brc{\zeta_{2}^6, \zeta_{2}^3\zeta_{3}^2, \zeta_{3}^4, \zeta_{2}^2\zeta_{3}\zeta_{5}, \zeta_{2}\zeta_{5}^2, \zeta_{2}^2\zeta_{5,3}, \zeta_{6,4,1,1}, \zeta_{2}\zeta_{3}\zeta_{7}, \zeta_{5}\zeta_{7}, \zeta_{2}\zeta_{7,3}, \zeta_{3}\zeta_{9}, \zeta_{9,3}}
    }.
    \label{eq:mzvbasis}
\end{align}
%

\section{Tensor perspective on the coproduct bootstrap}
\label{app:tensorCoprod}
Here we give another point of view on the integrability and Steinmann constraints discussed in \secref{ssec:coprods} using tensorial notation.
To generate a DS function $\stf\supbrk{\weight} \in \funspace_{\weight}$ at weight $\weight$ lets us form an ansatz for the
$\brc{\weight - 1, 1}$ coproduct component:
\begin{align}
    \cpr_{\weight - 1, 1}\brk{\stf\supbrk{\weight}} = \sum_{j, \alpha} c_{j \alpha} \, \stf\supbrk{\weight - 1}_j \otimes \log\brk{\alpha}
    \label{eq:cprF}
\end{align}
with unknown rational coefficients $c_{j \alpha} \in \Rationals^{\abs{J_{\weight - 1}} \times \abs{\alphabet}}$.
Next we apply eq. \eqref{eq:cpr1} to functions $\stf\supbrk{\weight - 1}_j$ of the previous weight, and introduce a set of integrability
and extended Steinmann conditions $\intSet$ on the $\brc{\weight - 2, 1, 1}$ component of the produced coproduct shown in eqs. \eqref{eq:integrability}
and \eqref{eq:steinmann}
\brk{see also the discussion around eq. \eqref{eq:intCondMap}}. For practical purposes, we collect these constraints in a single tensor
$\intTensor{\alpha \beta s} \in \Rationals^{\abs{\alphabet} \times \abs{\alphabet} \times \abs{\intSet}}$, whose non-zero elements are summarized below:
\begin{align}
    \intTensor{1, 1, 5} = - \intTensor{1, 2, 5} = \tfrac12 \intTensor{1, 3, 5} = \intTensor{1, 4, 5} = \intTensor{1, 5, 5} = 1, \nn\\
    \intTensor{2, 3, 4} = - \intTensor{2, 4, 3} = - \intTensor{2, 4, 5} = \intTensor{2, 5, 4} = 1, \nn\\
    - \intTensor{3, 3, 4} = - \intTensor{3, 3, 5} = \intTensor{3, 4, 3} = \intTensor{3, 5, 3} = 1,
    \label{eq:intTensorDef}\\
    \intTensor{4, 1, 5} = - \intTensor{4, 2, 5} = - \intTensor{4, 3, 4} = \intTensor{4, 4, 3} = - \intTensor{4, 5, 1} = \intTensor{4, 5, 2} = 1, \nn\\
    - \intTensor{5, 1, 4} = \intTensor{5, 4, 1} =
    \intTensor{6, 3, 2} =
    \intTensor{7, 2, 3} =
    - \intTensor{8, 1, 2} = \intTensor{8, 2, 1} = 1. \nn
\end{align}
The action of these conditions on the $\brc{\weight - 2, 1, 1}$ coproduct component of the unknown function $\stf\supbrk{\weight}$ follow from eq.
\eqref{eq:cprF} and read:
\begin{align}
    \sum_{\substack{j, k \\ \alpha, \beta}}
    c_{j \alpha}
    \cc{\weight - 1}{j k \beta} \,
    \stf\supbrk{\weight - 2}_{k} \,
    \intTensor{s \alpha \beta}
    =
    \sum_{\substack{j, k \\ \alpha}}
    \stf\supbrk{\weight - 2}_{k} \,
    M_{k s j \alpha} \,
    c_{j \alpha}
    = 0, \quad \text{for each } s \in \intSet,
    \label{eq:intCond}
\end{align}
where
$M_{k s j \alpha} \in \Rationals^{\abs{J_{\weight - 2}} \times \abs{\intSet} \times \abs{J_{\weight - 1}} \times \abs{\alphabet}}$ \brk{implying the necessary
transposition of tensor indices here} was already defined in eq. \eqref{eq:Mdef} above.
Now note, that the DS functions of lower weight $\stf\supbrk{\weight - 2}_k$ are linearly independent from one another by construction, which means that eq. \eqref{eq:intCond}
contains $\abs{J_{\weight - 2}} \cdot \abs{\intSet}$ constraints for $\abs{J_{\weight - 1}} \cdot \abs{\alphabet}$ unknowns.  The space of DS functions
at weight $\weight$ lies inside the nullspace of the
$M_{\brk{k s} \brk{j \alpha}} \in \Rationals^{\brk{\abs{J_{\weight - 2}} \cdot \abs{\intSet}} \times \brk{\abs{J_{\weight - 1}} \cdot \abs{\alphabet}}}$
matrix: schematically we write
$\funspace_{\weight} \subset \kernel\brk{M_{\brk{k s} \brk{j \alpha}}}$, where the brackets $\brk{k s}$ and $\brk{j \alpha}$ denote vectorization of
indices. In \mathematica{} such a vectorization $M_{k s j \alpha} \to M_{\brk{k s} \brk{j \alpha}}$ of a rank 4 tensor into a rank 2 one is achieved
simply via application of \code{Flatten[\#, \{\{1, 2\}, \{3, 4\}\}]\&}. The further processing of this nullspace is described in the main text in
\secref{ssec:coprods}.

\bibliographystyle{nb}
\bibliography{references}

\end{document}